\documentclass{article}
\usepackage{blindtext}
\usepackage[a4paper, total={6.5in, 10.8in}]{geometry}
\usepackage{graphicx} % Required for inserting images
\usepackage{cite}
\usepackage{caption}
\usepackage{subcaption}
\usepackage{amsmath}
\usepackage{amssymb}
\usepackage{amsthm}
\usepackage{soul}
\usepackage{placeins}
\usepackage{comment}
\usepackage{xcolor, float}
\usepackage{authblk}
\usepackage{hyperref}
\usepackage{tabularray}
\usepackage{textcomp}
\usepackage{gensymb}
\newcommand{\mycomment}[1]{}
\usepackage{xr}
\usepackage{cleveref}

\makeatletter
\newcommand*{\addFileDependency}[1]{% argument=file name and extension
\typeout{(#1)}% latexmk will find this if $recorder=0
\@addtofilelist{#1}
\IfFileExists{#1}{}{\typeout{No file #1.}}
}\makeatother

\newcommand*{\myexternaldocument}[1]{%
\externaldocument{#1}%
\addFileDependency{#1.tex}%
\addFileDependency{#1.aux}%
}

\myexternaldocument{supplementary_materials_submission}

\begin{document}

\title{A perspective on protein structure prediction using quantum computers}

\author[1]{Hakan Doga}
\author[2]{Bryan Raubenolt}
\author[2]{Fabio Cumbo}
\author[2]{Jayadev Joshi}
\author[2]{Frank P. DiFilippo}
\author[2]{Jun Qin}
\author[2]{Daniel Blankenberg}
\author[3]{Omar Shehab}

\affil[1]{IBM Quantum, Almaden Research Center, San Jose, California 95120, USA}
\affil[2]{Center for Computational Life Sciences, Lerner Research Institute, The Cleveland Clinic, Cleveland, Ohio 44106, USA}
\affil[3]{IBM Quantum, IBM Thomas J Watson Research Center, Yorktown Heights, NY 10598, USA}
\date{}

\maketitle

\begin{abstract} Despite the recent advancements by deep learning methods such as AlphaFold2, \textit{in silico} protein structure prediction remains a challenging problem in biomedical research. With the rapid evolution of quantum computing, it is natural to ask  whether quantum computers can offer some meaningful benefits for approaching this problem. Yet, identifying specific problem instances amenable to quantum advantage, and estimating quantum resources required are equally challenging tasks. Here, we share our perspective on how to create a framework for systematically selecting protein structure prediction problems that are amenable for quantum advantage, and estimate quantum resources  for such problems on a utility-scale quantum computer. As a proof-of-concept, we validate our problem selection framework by accurately predicting the structure of a catalytic loop of the Zika Virus NS3 Helicase, on quantum hardware.
\end{abstract}

\section{Introduction}
\begin{figure}[h!]
    \centering
    \includegraphics[scale=0.5]{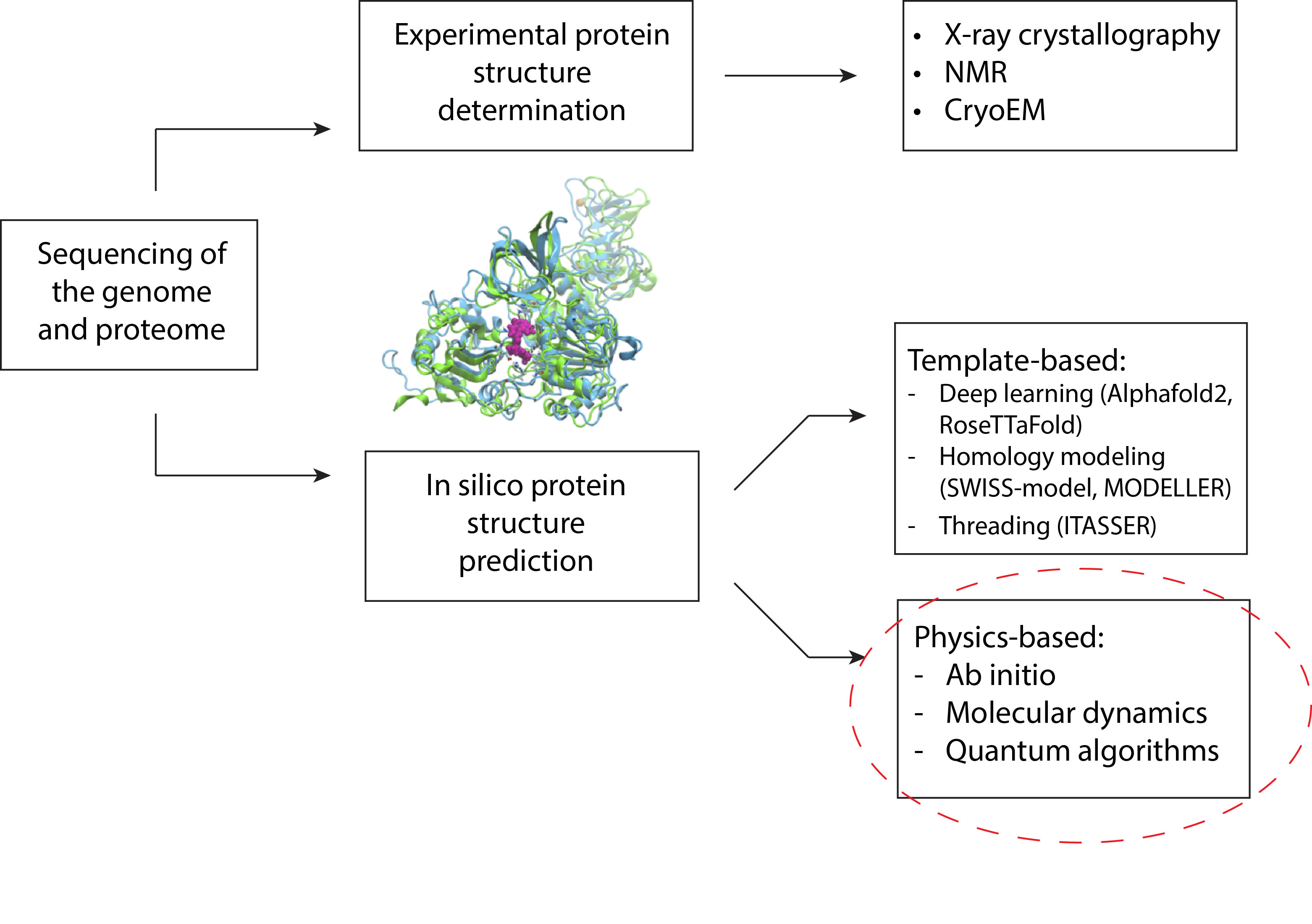}
    \caption{ \textit{Overview of the PSP pipeline. Following genomic sequencing, the primary amino acid sequence is determined. The experimental method then starts with expressing this protein by genetically modifying another organism with this new sequence. This organism will then translate these proteins, and the new protein of interest can be isolated, purified, and then solved using X-ray crystallography, NMR, or CryoEM. The \textit{in silico} methods on the other hand, simply take the primary amino acid sequence as input and the structure is predicted by either a physics-based method (where the underlying biophysics is somehow simulated) or a template-based method (where machine learning algorithms predict structures based on patterns found in a training set of experimental templates). The method we adopt in this work falls under the category of physics-based algorithms. As an illustrated example, an in silico model and X-ray crystal structure of the SARS-CoV2 NSP13 helicase (PDB: 7NN0) are superimposed, along with a docked known inhibitor (colored in magenta)}.}
    \label{tc10b_tc5b}
\end{figure}

The intricate dance of life at the molecular level is orchestrated by proteins, with virtually all biological activity tied to the three-dimensional conformations they adopt\footnote{Corresponding authors: Hakan Doga (hakandoga@ibm.com) and Bryan Raubenolt  (raubenb@ccf.org).}. The phenomenon by which these structures are predetermined from their primary amino acid sequence is known as the ``protein folding problem", and it is inherently central to all life and its myriad of diseases. In nature, protein folding is a path-dependent process, meaning that the optimal path is usually taken \cite{ken_dill_psp, kend_dill_psp2}. For most biomedical research applications, predicting the optimal structure itself, without necessarily reproducing the optimal path, is arguably most important and more attainable. This is  known as protein structure prediction (PSP) (see Figure ~\ref{tc10b_tc5b}). Decades of scientific inquiry have sought to unravel the mysteries of how and why proteins assume their complex structures, often driven by the quest for understanding diseases at the molecular level. To comprehend the function of a protein and devise effective drugs targeting them, an accurate depiction of their physiologically active structure is indispensable. Traditionally, this has been achieved through laborious wet lab experiments involving genetic modifications, protein isolation, and purification. Techniques like X-ray crystallography, NMR, and CryoEM have been instrumental in solving protein structures, revolutionizing our grasp of diseases. However, these methods are time-consuming, expensive, and not without limitations. Recognizing the need for alternatives, researchers turned to machine learning, exemplified by AlphaFold2  \cite{Jumper2021}, RoseTTaFold \cite{Baek2021}, and ITASSER \cite{Zhou2022}, which leverage experimentally determined structures. While transformative, these methods may lack a nuanced understanding of the underlying physics, potentially hindering predictions of novel protein structures \cite{Outeiral2022}. Physics-based methods on the other had, such as molecular dynamics (MD) simulations, face challenges in scalability and practicality. Interested readers may refer to \cite{noe2020machine, kuhlman2019advances} for comparative analyses of the above-mentioned methods.

Fundamentally, protein folding is a physics problem. A computer algorithm armed with a realistic thermodynamic description and mathematical framework for modeling interactions between amino acids (residues) could theoretically navigate the vast conformational space, ultimately arriving at an optimal solution \cite{kend_dill_pf_limit}. Yet, the computational complexity of this challenge has made it a persistent hurdle in the life sciences. In this landscape, quantum computing holds the potential to provide meaningful utility for this problem.  While there have been major advances in quantum hardware and algorithm development \cite{bravyi2022future}, finding the appropriate class of problems amenable to quantum advantage is still an open question for all areas of practical applications. Estimating the resources needed to enable quantum advantage is equally challenging, with sporadic progresses. In this work, we provide our perspective, a scalable framework to identify PSP problem instances that may be amenable to quantum advantage. Narrowing the problem space through a combination of protein sequence length, disorder, number of mutations and multiple sequence alignment (MSA) depth, we opine on how to select problems at different scales that have been found challenging for state-of-the-art classical methods. Furthermore, this approach makes our problem selection framework explainable.

Section \ref{qc-survey} introduces quantum computing fundamentals, laying the groundwork for understanding its application in solving the protein folding problem. Section \ref{main_section} describes  our perspective. It delves into the complexities of PSP, highlighting challenges related to protein size, mutations, and the role of MSAs. The focus is on identifying the problem subspace where quantum computing might outperform deep learning methods reliant on MSAs. Experimental results in Section \ref{results-workflow} showcase our quantum-classical hybrid workflow predicting the structure of the seven amino acid catalytic P-loop from a vital Zika virus protein (NS3 Helicase) using the IBM\_Cleveland quantum computer, validating our perspective with a concrete example. Section \ref{resource_est} discusses the scalability of the quantum algorithm and provides high level quantum resource estimation for PSP problems. The conclusion reflects on the significance of the work, opening avenues for future exploration in leveraging quantum computing for protein folding challenges.

\section{Quantum computing: A  brief introduction}\label{qc-survey}
Quantum Computing is a new model of high performance computing where the traditional foundation of computing, i.e. binary logic, has been replaced by theories of quantum mechanics \cite{dirac1981principles, heisenberg1973development}. This section aims to provide a very concise introduction to the topic, while the readers are encouraged to review \cite{nielsen2010quantum} for a comprehensive overview, and Section 2 of \cite{basu2023towards} for a primer from the perspective of healthcare and life sciences.  The power of quantum computing comes from quantum mechanical effects such as superposition, entanglement, negative state probability i.e. interference, and probabilistic measurement. These phenomena sometimes allow a quantum algorithm to naturally map the degrees of freedom of quantum hardware to those of a target quantum system and simulate in an efficient manner. For some suitably structured problems, it is possible to exploit these phenomena to design quantum algorithms capable of traversing a search space or optimization cost landscape in a more efficient probabilistic manner. Over the last few decades, quantum advantages have been theoretically demonstrated for prime factorization \cite{shor1994algorithms}, unstructured search \cite{grover1996fast}, network flows  \cite{ambainis2006quantum}, quantum simulation \cite{lloyd1996universal}, topological invariants \cite{aharonov2006polynomial}, partition functions \cite{wocjan2008quantum},  semidefinite programming \cite{kalev2019quantum},  linear systems \cite{harrow2009quantum},  differential equations \cite{engel2019quantum}, dynamic programming \cite{ambainis2019quantum}, bilinear functions \cite{o2023quadratic}, etc. A few recent works have also reported quantum utility \cite{kim2023evidence} or empirical quantum advantage \cite{daley2022practical, riste2017demonstration, o2023quadratic, dale2015provable, stamatopoulos2022towards, krunic2022quantum, wu2021application}, as defined in \cite{krunic2022quantum}. Several architectures have been proposed for scalable quantum computers including neutral atom qubits \cite{levine2019parallel}, spin qubits \cite{burkard2021semiconductor}, topological qubits \cite{conlon2019error}, trapped-ion qubits \cite{kozhanov2023next}, and superconducting qubits \cite{steffen2011quantum}. With 30+ quantum computers, IBM Quantum Experience is the first ever and largest cloud-based quantum computing service. Quantum software development kits like Qiskit \cite{cross2018ibm}, CUDA Quantum \cite{the_cuda_quantum_development_team_2023_8175982}, Forest \cite{computing2019pyquil}, PennyLane \cite{bergholm2018pennylane}, CirQ \cite{hancockcirq} and Braket \cite{gonzalez2021cloud} are also available with increasingly more application verticals in each release. Several vendors have also announced their developmental road maps \cite{IonQ, Quantinuum, bourassa2021blueprint, bartolucci2023fusion} with IBM leading the way to achieve one hundred thousand qubits with increased gate quality and speed within a decade \cite{bravyi2022future, IBM_Quantum_Roadmap_2015}. In parallel with the progress of technologies, the research communities have also presented their point of views for using quantum computers in several areas of applications including scientific discovery \cite{alexeev2021quantum}, biological sciences \cite{emani2021quantum}, nuclear physics \cite{beck2023quantum}, high-energy physics \cite{di2023quantum}, cell-centric therapeutics \cite{basu2023towards}, financial engineering \cite{egger2020quantum}, climate science \cite{berger2021quantum}, etc. Finally, the responsible usage of quantum computing is also emerging as an area of research \cite{kop2023towards}.

\subsection{Quantum search and optimization algorithms}
\label{quantum_algos}
While finding the global minima of the  energy landscape of protein conformation is an optimization problem, one could also imagine it as a search problem. In that case, the database consists of all possible conformational energies with the lowest energy as the marked item to search for. The quantum algorithm for unstructured database search is Grover's algorithm which provides quadratic speedup over state-of-the-art classical search algorithms \cite{grover1996fast}. The algorithm initializes a quantum state in an equal superposition over all database entries. It then iterates a step which amplifies the amplitude of the target entry. After  $O(\sqrt{N})$ iterations, a measurement will find the entry with high probability, and it has been shown as optimal \cite{bennett1997strengths}. The optimality of Grover's algorithm is also supported by the observation that it defines a geodesic in quantum Hilbert space \cite{alvarez1999comment, cafaro2012grover}. Therefore, any other quantum search or optimization algorithm may be considered as a parametrized approximation of Grover's algorithm. One such algorithm is quantum approximate optimization algorithm (QAOA) \cite{farhi2014quantum}.  It works by setting up a cost Hamiltonian whose ground state encodes the search result. A quantum system is initialized in an easy-to-prepare state. A series of unitary operations alternately apply a mixer Hamiltonian and the cost Hamiltonian. At the end, the state of the system is measured, giving a candidate solution. While low-depth QAOA is not expected to outperform the state-of-the-art classical algorithm\cite{bravyi2020obstacles, hastings2019classical} (see \cite{de2023limitations, clary2023exploring} for a more formal study of the limitations of variational quantum algorithms), at a higher depth the algorithm increasingly becomes an approximation of the Grover's algorithm and has the potential to maximize fractional Grover speedup \cite{farhi2014quantum}. See Section ~\ref{supp-sec:quantum} of the supplementary materials for a detailed discussion of algorithmic structures needed for quantum speedup.

\subsection{The protein folding funnel and the prospect of quantum advantage}
\label{funnel-advantage}
\begin{figure}[h!]
    \centering
    \includegraphics[scale=0.4]{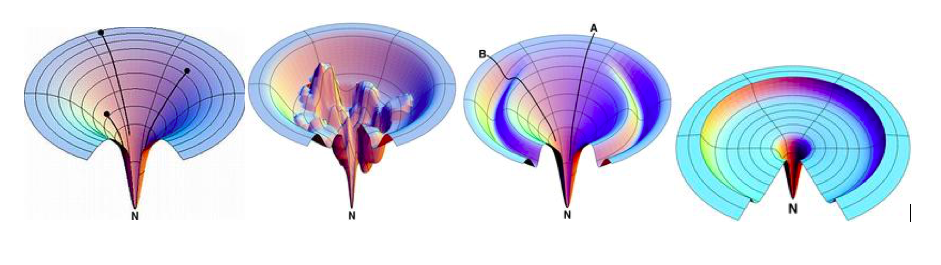}
    \caption{This graphic was originally created and released in the public domain by Ken A. Dill. Original caption: \textit{The illustrations of proposed energy landscape that each demonstrate the degree of freedom a protein possesses in terms of configurations and the multidimensional routes that a protein can take to achieve its final configuration. From left to right for proposed funnel-shaped energy landscape: the idealized smooth funnel, the rugged funnel, the Moat funnel, and the Champagne Glass funnel \cite{FoldingFunnel}}.}
    \label{fig:funnels}
\end{figure}

The theory of protein folding considers how proteins' primary amino acid sequences dictate how they fold rapidly and specifically into their native 3D structures. The classical view was that folding occurred through discrete intermediates along a linear pathway. In contrast, the energy landscape theory views folding as a progressive organization of an ensemble of partially folded structures funneled towards the native state \cite{dill1989thermal, dill1997levinthal, onuchic2004theory}. Evolution has shaped proteins to have a rugged, funnel-like landscape biased towards the functional native structure.  The funnel shape implies folding robustness, with different routes down the funnel possible for the same protein. Just like most other spontaneous thermodynamic phenomena, protein folding is largely path dependent - nature takes the optimal path to the optimal solution. The native state's stability increases moving down the funnel through local folding events. Common patterns emerge like topology determining mechanisms, but details depend on subtle sequence variations. The funnel concept explains how folding can be fast despite the vast number of possible non-native conformations. The landscape guides the protein through a directed search. Figure ~\ref{fig:funnels} shows four proposed folding funnels.

Finding the global minimum energy in the folding funnel can be understood as a search problem where the database entries are conformational energies. The funnel shape indicates that despite the overall ``easy-to-follow" macro structure, the ruggedness on the funnel wall and the bottom induces hardness to conformation prediction. 

The ruggedness of the energy landscape at the bottom of the folding funnel correlates with the complexity and diversity of conformations a protein can adopt \cite{ma1999folding, clark2004protein, wolynes1995navigating, karplus2011behind, onuchic1995toward, tsai1999folding, chikenji2006shaping, jin2003novo, hansmann1999folding, onuchic2004theory, socci1997exploring}. A smooth, single-minimum funnel bottom indicates a rigid protein with few accessible conformers. A rugged funnel bottom with many minima reflects a flexible protein existing as an ensemble of diverse conformers interconverting between sub-states. The more rugged the bottom, the more conformational flexibility the protein has to sample alternate structures. Proteins with rugged, multi-minima landscapes can bind ligands in a non-specific way by selecting conformers complementary to each ligand from their ensemble. In contrast, proteins with smooth funnels have less conformational heterogeneity and exhibit more specific binding to pre-organized native structures. Increased ruggedness also enables phenomena like induced fit binding, crystal packing effects, domain swapping, and misfolding aggregation by allowing access to alternate conformers. Molecular chaperones may smooth rugged landscapes to promote proper folding over misfolding. In summary, the complexity and interconversion of a protein's conformational sub-states is directly related to the ruggedness of the energy landscape at the bottom of its folding funnel \cite{tsai1999folding}. Rugged landscapes confer flexibility and multi-specificity, while smooth funnels restrict accessible conformers and interactions.

The ruggedness in the folding funnel has a curious parallel in unstructured search problems when one interprets the problem as traversing the geodesic of a search space (Grover's quantum search algorithm) or optimizing a rugged cost function using a variational algorithm with a parameterized quantum circuit (QAOA algorithm). Besides, the space of the folding pathways has a natural tree structure (See Figure 5 of \cite{onuchic1997theory}) similar to the search tree structure in Grover solution space. These algorithms have been briefly introduced in Section ~\ref{quantum_algos}. It has been observed that the ruggedness of the search or optimization landscape creates amenability for quantum advantage \cite{bapat2018bang, wu2011role, baldwin2018quantum, leier2003exploring, stepney2008searching, streif2020training, mcclean2021low, wang2023quantum, wang2020x, shaydulin2023parameter} (see Section ~\ref{supp-sec:quantum} of the supplementary materials for a detailed discussion). The quantum advantage appears when there is a least amount of information or a highest amount of uncertainty about the search space which is also known as the ``worst case" in algorithmic complexity analysis. In the case of Grover's algorithm one can attribute this quantum speed up to the fact that quantum mechanics allow us to associate negative probabilities to search paths. Therefore, a subset of the search paths will cancel each other, rendering the search problem smaller and easier. If one is using the QAOA algorithm, the same phenomena may be viewed as an effect of quantum tunneling where a quantum search process may cross energy barriers in a probabilistic manner that does not have any classical analog. Under these circumstances, we can conjecture that the more rugged a folding funnel is, the easier it will be for a quantum algorithm to find the lowest energy conformation relative to the performance of its classical competitors.

\section{What makes protein structure prediction hard?}\label{main_section}

\subsection{Performance of physics-based methods against increasing sequence length}
\begin{figure}[h!]
    \centering
    \includegraphics[scale=0.6]{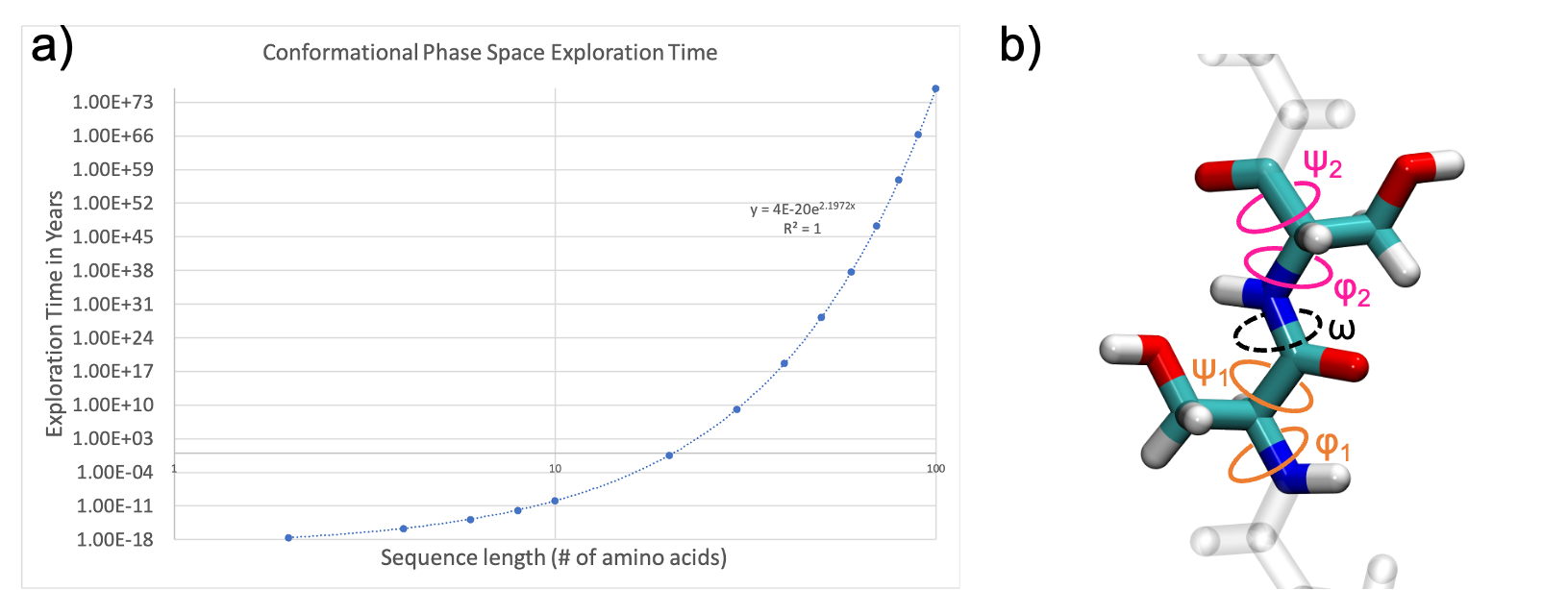}
    \caption{\textit{. A graphical representation of the Levinthal paradox a) The data considers that for a protein with \textit{n} amino acids, there are \textit{n}-1 peptide bonds. For each peptide bond, there are also 2 other bond angles on either side of the $\alpha$ carbon, $\phi$ and $\psi$, and assumes that each peptide in the sequence can adopt up to 3 conformations (3 combinations of these bond angles). b) So, for a protein with \textit{n} amino acids, there are a total of 3$^{2(n-1)}$ possible conformations. Assuming 1 picosecond is spent to sample each conformation, the y-axis represents the total exploration time in years to sample all possible conformations for a protein with \textit{n} amino acids.}}
    \label{levinthal}
\end{figure}

Regardless of the physics-based computational method used, sequence length quickly becomes a major limitation. As a protein sequence becomes larger, there is an exponential increase in the search space (number of possible conformations), and a corresponding exponential growth in the required run time for an exhaustive search (see Figure \ref{levinthal}a). Most non-molecular dynamics (non-MD) PSP methods (also known as \textit{ab initio} or free modeling methods) have generally been limited to structures of only a few dozen amino acids in length. One study by M. Yousef et al \cite{Yousef2019} compared the performance of four of these methods. While most of these methods appeared to produce backbone Root Mean Squared Deviations (RMSDs) within three angstroms when compared to the experimental results, the investigated structures were all 31 amino acids or less in length. PEP-FOLD3 \cite{pepfold3} is designed to model peptides between 5 and 50 amino acids, and has been successful in doing so in different studies. Quark \cite{quark}, an \textit{ab initio} program developed by the same group who created ITASSER \cite{itasser}, can predict structures accurately, but was mainly designed for fragments up to 20 amino acids in length. For small peptides at this scale, these programs perform well (run times of less than 14 hours for all test cases reported in \cite{Yousef2019}) and can offer viable solutions with the right compute resources. For larger, biologically relevant proteins, physics-based classical algorithms like these may not be as suitable due to the rapid increase in the conformational space.

Although there has been remarkable success in some studies where all-atom MD simulations are applied to predict folded structures, with the most impressive results arguably being those of D.E. Shaw's research group \cite{anton1, anton3}, sequence length still remains a limitation. These efforts led to the simulation of possible folding paths, as well as an accurate prediction of the folded state, for human ubiquitin \cite{anton2} and a set of 12 ``fast folding" proteins \cite{anton4}. The main issue is the amount of sampling time (the length of the simulation in time, as defined by the product of the total number of discrete time steps and step size) as well as compute resources it took to achieve these results, since most researchers may have limited access to either. Although the majority of the structures in \cite{anton4} folded to near perfect accuracy, much of this occurred in the order of hundreds of microseconds to milliseconds of sampling time (depending on the system size, this generally requires significant run times ranging from weeks to months even when using conventional HPC resources) and the investigated proteins were relatively small, ranging in size from roughly 10-80 residues. Much like the other physics-based methods, standard (unbiased) all-atom molecular simulations still present a threshold in sequence size for which they may no longer be a practical solution for predicting larger protein structures. One alternative for speed up is of course MD simulations using coarse grain potentials, in particular those derived from neural networks \cite{Majewski2023}, but among the limitations is the fact that long time scale all-atom simulations still need to be run for the neural network to train on, and even then these potentials are still not generalizable (they cannot be readily used for proteins largely different from the training set).

This gap between practical sampling times in simulation and the real folding times in experiments has been known for quite some time, and it is one of the main reasons enhanced sampling methods (such as replica exchange molecular dynamics REMD) were developed. REMD is an excellent tool for thermodynamic sampling of large conformational spaces, with relatively short time scales. Rather than having a single, long simulation of the system, several ``replicas" of the same system are simulated at different temperatures (usually room temperature, as well as above and below the melting points) and different random seeds for the initial velocities. In most implementations, the idea is that these temperature-dependent simulations can then undergo Monte Carlo swaps between neighboring replicas, preventing the lower temperature conformations from becoming ``trapped" in a local minima. While the sampling time can be orders of magnitude less than running a single conventional simulation at fixed temperature, that is compensated by the number of simulations one has to run, which can regularly exceed a few dozen replicas even for a small 20 amino acid protein like Trp-Cage \cite{trp-cage-remd2, trp-cage-remd3}. Because of this burden, new methods of REMD have been developed which can significantly reduce the number of replicas required to sample the same conformational space \cite{reds}.

\subsection{Mutations and intrinsically disordered proteins}
% \textcolor{red}{Shehab: SHOULD WE FURTHER ORGANIZE AS 3.2.1 MUTATIONS AND 3.2.2 INTRINSICALLY DISORDERED PROTEINS? Hakan: I think it looks okay at the moment and the discussion is self-contained, but we can discuss this after we receive some feedback internally and when we want to submit it to journal.} 

One of the main advantages of the template-based deep learning methods like AlphaFold2 and RoseTTaFold is the sheer size of the structures they are able to produce. They are not limited to a few dozen residues as we've seen with the physics-based methods. Both AlphaFold2 and RoseTTaFold can readily produce models up to a couple thousand residues in length, in part due to the fact that their data bases (PDB\footnote{https://www.rcsb.org/.}) include experimentally determined structures which span across this size range. Obviously the larger the sequence, the more MSAs and fragments are needed, and thus computing power. The upperbound is roughly between 1280 and 2700 amino acids, according to the European Bioinformatics Institute \footnote{https://alphafold.ebi.ac.uk/faq.}. Despite this impressive capability, success still depends on how accurate the models are.  Specifically, the models are known to produce discrepancies when dealing with proteins with a) mutated sequences and b) intrinsically disordered regions. Here we discuss the effect of these two factors. 

\subsubsection{Mutations}

While the ability for these programs to predict wildtype proteins containing mostly ordered secondary structures, i.e. localized structures that form based on interactions within the protein backbone, is unquestionable, significant errors have been reported when attempting to predict mutated variants. The discrepancies can be observed in both the 3D coordinates of the produced structure as well as the predicted thermodynamic stability of the mutated residues relative to their confidence scores \cite{af2_mutations, af2_mutations2}. Mutations change the local chemical space, and thus the physics of the interacting residues. Even the slightest mutations between similar residues (for example valine to leucine) can have notable impacts on the local structure. Some mutations can be drastically different, such as going from a slender, flexible, charged residue like glutamic acid to a bulky hydrophobic sidechain like tryptophan. Such cases can have profound effects, such as disrupting charge distributions beyond the local structure and often breaking relevant hydrogen bond networks vital to the protein's activity. In order for PSP methods to be successful, the algorithm needs to capture the consequences of mutations. From a global perspective, mutations can change the free energy landscape of a protein. This can sometimes lead to a more rugged landscape (or more smooth), perhaps introducing new local minima (see Section \ref{funnel-advantage} ). This can be problematic, as an algorithm's optimization protocol could get ``stuck" in these regions, producing a conformation it thinks corresponds to the lowest energy, particularly if heuristic optimizers are involved. Of course exhaustive search algorithms could naturally avoid this, but they are not scalable beyond a certain size. 

\begin{figure}[h!]
    \centering
    \includegraphics[width=\textwidth
    ]{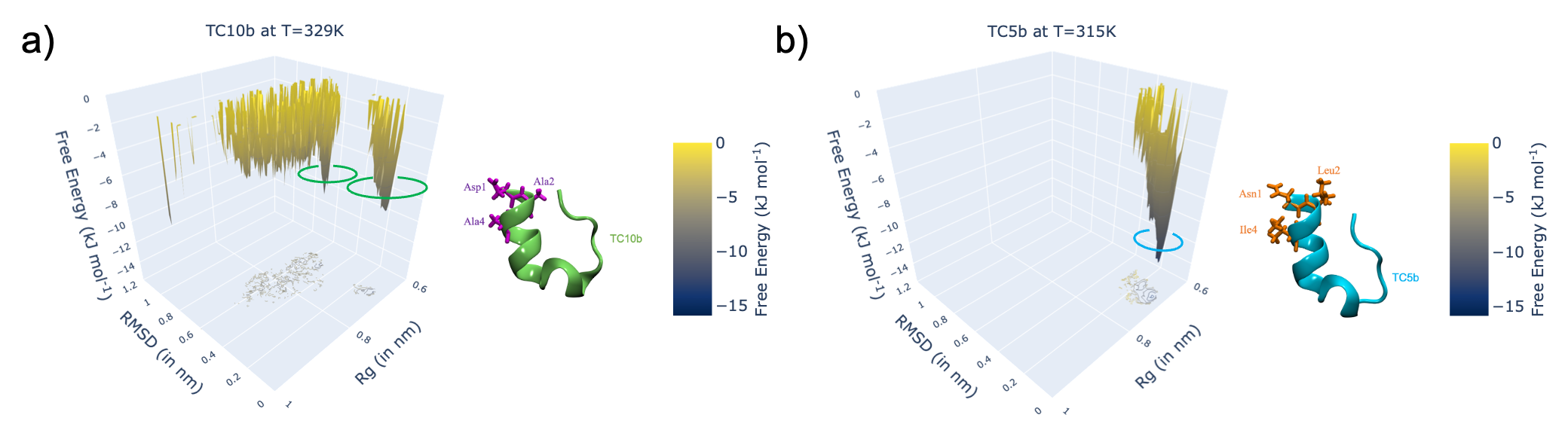}
    \caption{\textit{Free energy landscapes of a) TC10b and b) TC5b derived from molecular dynamics simulations at their melting points of 329K and 315K, respectively. The folded NMR structures are quite similar (0.8 \r{A} RMSD between them), but the free energy landscape is dramatically different, with TC10b demonstrating two possible folding pathways and their minima, as opposed to the more well defined funnel and obvious global minimum in the case of TC5b. The three point mutations between these structures give rise to different physicochemical properties such as their melting temperatures, and accordingly different free energy landscapes.}}
    \label{tc10b_tc5b_energy}
\end{figure}

The Trp-cage proteins are 20 amino acid engineered ``miniproteins" that have long been studied as ideal benchmarking candidates for physics-based PSP methods \cite{trp-cage-remd, trp-cage-remd2,trp-cage-remd3,md4psp,anton4}. One of the original variants of this protein is TC10b (PDB: 2JOF), with sequence DAYAQWLKDGGPSSGRPPPS. A closely related variant is TC5b (PDB: 1L2Y), with sequence NLYIQWLKDGGPSSGRPPPS. The structures share a sequence identity of 85\%, with mutations occurring at residues 1, 2, and 4. The mutations at 2 and 4 still preserve the hydrophobic chemical space, but result in more bulky sidechains (leucine and isoleucine instead of alanines). The mutation at 1 is more pronounced however, with a charged aspartate residue being substituted by an uncharged, but still polar, asparagine residue. Although the NMR structures are quite similar, with an RMSD of 0.8 \r{A} between the two, the folding pathways and free energy landscapes can be noticeably different. 

Proteins at their melting point are in a state of unfolding and refolding, much like a physical system transitioning between two states. One way to model what this looks like is through molecular dynamics simulations. As a proof of concept, we can explore the free energy landscape of the two variants of Trp-Cage, TC10b and TC5b \cite{trp-cage2008}, derived from conventional molecular dynamics simulations at their theoretical melting points (329K and 315K, respectively). These simulations were performed with GROMACS2022 \cite{gromacs} (as implemented in the Galaxy platform \cite{galaxy}) using the FF99SB force field \cite{Hornak2006} and the TIP3P water model \cite{tip3p}, for approximately 500 nanoseconds in the isothermal-isobaric (NPT) ensemble. Time-dependent RMSDs and radii of gyration (R$_g$) of the alpha carbon backbone were extracted to measure conformational changes over time, and then posteriorly analyzed using MD DaVis \cite{dibyajyoti_maity_2022_6227047}. Figure \ref{tc10b_tc5b_energy} illustrates the resulting data, where the free energy (z-axis) is plotted as a function of RMSD (x-axis) and R$_g$ (y-axis). The difference between both plots is quite clear. In Figure ~\ref{tc10b_tc5b_energy}a, TC10b presents a much more rugged landscape, so much so that there appears to be almost two separate folding paths, leading to two distinct optimal conformations (circled in green). Both minima are relatively close, with the smaller funnel reaching -9.01 kJ/mol and the larger funnel arriving at -9.4 kJ/mol. Both optimal conformations appear to share nearly the same Rg (6.8 vs 6.9 \r{A}), while having distinct RMSDs (0.8 vs 5 \r{A}). TC5b on the other hand, presents a much more conservative free energy landscape (Figure \ref{tc10b_tc5b_energy}b). There is a single global minima (circled in cyan) corresponding to an energy of -13.5 kJ/mol. This optimal conformation shares the same radius of gyration as TC10b (6.9 \r{A}) with a slightly different RMSD (0.7 \r{A}). Thus, assuming the force fields are reproducing the experimental melting points accurately (which is beyond the scope of this work) this brief qualitative exercise highlights a central thesis of this perspective: even when mutated sequences still lead to similar experimentally solved structures, the free energy landscape and dynamic folding pathways can be quite different. Computational methods must therefore be aware and capable of capturing these unique interactions, while enumerating and sampling the entire conformational space. This is precisely what a quantum algorithm could be capable of doing, at scale.

\subsubsection{Intrinsically disordered regions}
While most proteins have ordered domains (with $\alpha$-helices, $\beta$-sheets, and $\beta$-barrels), many intrinsically disordered regions (IDRs) can exist throughout the structure. In fact, there are many proteins that are significantly disordered (containing at least 30\% disordered residues), appropriately regarded as intrinsically disordered proteins (IDPs). The human proteome is estimated to be approximately 32\% intrinsically disordered \cite{Deiana2019}. These regions serve a purpose as they are naturally flexible, allowing them to often serve as binding domains for other proteins or ligands. From a thermodynamic perspective, one can think of these regions as occupying a series of local minima in the free energy landscape, which are close in magnitude and thus equiprobable, with the global minimum often induced by a binding process. Despite their disorder, there is likely an ensemble average of structures, and it is important for PSP methods (and subsequent drug design efforts) to capture this accurately. The deep learning methods often fail at properly predicting these conformations \cite{idr1}. What appears to be consistent though, at least in the case of AlphaFold2, is that these regions are usually predicted with low confidence scores, and for those IDRs which conditionally fold (when bound to a substrate), a recent study determined that it usually predicts a conformation closely related to that folded state \cite{idr3}. This perhaps presents an unfavorable bias in the neural network, which could mislead researchers who are assuming the model is predicting the unbound protein. Another study found interesting correlations between the AlphaFold2 model's confidence scores and the observed flexibility throughout the protein in molecular dynamics simulations \cite{idr2}. Regions with lower confidence scores in the model were found to have higher flexibilities in the simulations, while those with higher scores usually corresponded to more stable secondary structures. This was quantified by measuring root mean square fluctuations (RMSFs), a common metric to assess flexibility in MD simulations and NMR ensembles. While this is an interesting correlation, this inference does not indicate that the predicted conformation of these IDRs is accurate, and researchers should be aware of this. To get a proper representation of these disordered regions, an ensemble average must be sampled. In order to do this, the physics at play must somehow be simulated, which is something the template-based methods remain arguably agnostic to \cite{Outeiral2022}.

\begin{figure}[h!]
    \centering
    \includegraphics[scale=0.5]{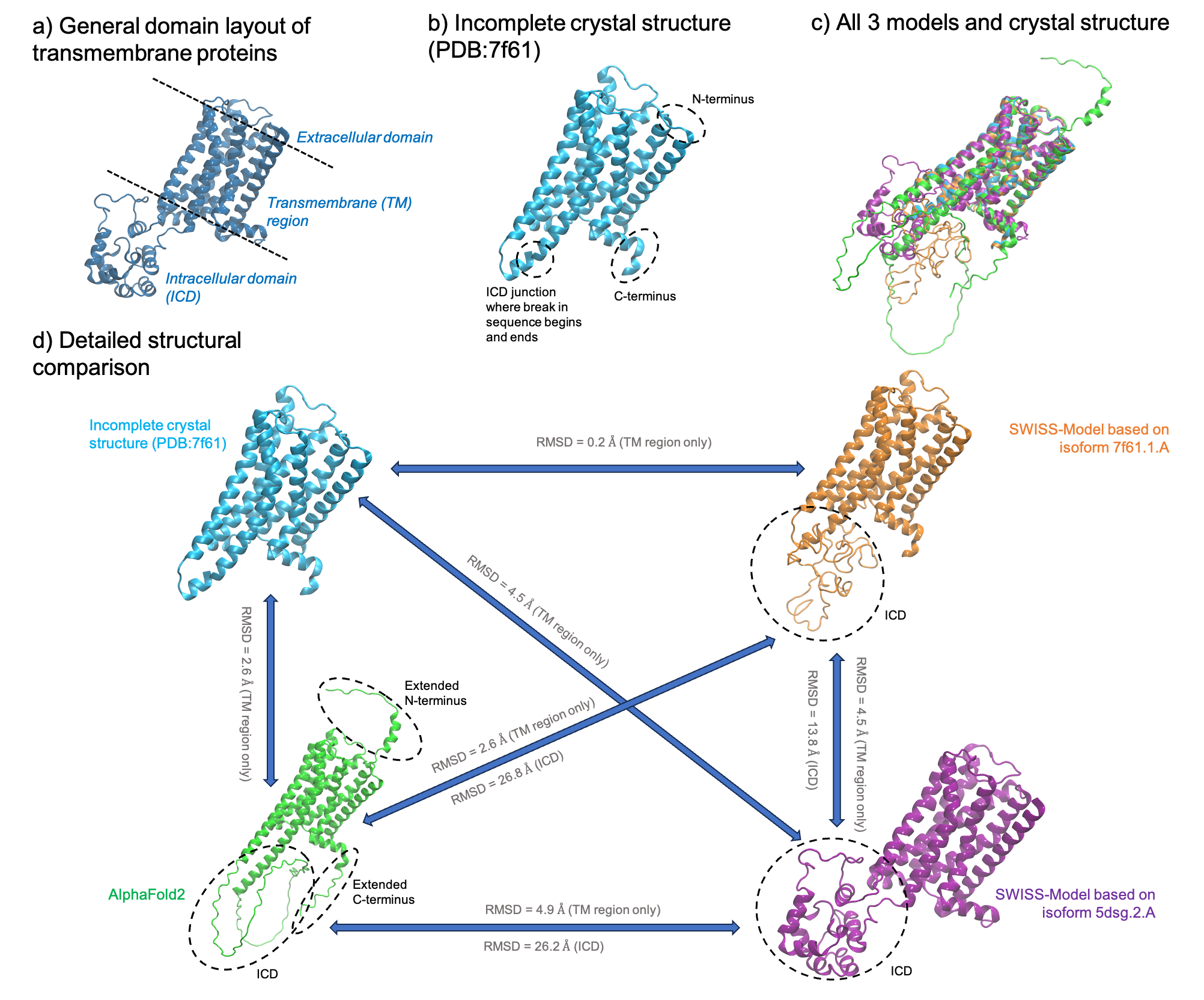}
    \caption{\textit{a) The common anatomy of transmembrane proteins like the G-protein coupled receptor (GPCR) superfamily. b) The incomplete experimentally determined crystal structure of H3HR, which is essentially missing the entire ICD. c) Alignment and comparison between the experimental structure and computational models of H3HR, highlighting the excellent agreement in the TM region, but the significant discrepancies in the predicted ICD. d) Detailed comparison of the structures reveals that the TM region is closely modeled with RMSDs of 0.2 to 4.9 \r{A}, while the ICD varies drastically with RMSDs of 13.8 to 26.8 \r{A}. The least accurate model is arguably AlphaFold2's, with a highly disordered ICD and an extended C- and N-terminus.}}.
    \label{hrh3_all}
\end{figure}

A perfect example of the importance of representing the disordered regions correctly occurs with transmembrane proteins. As the name implies, these proteins are embedded in cell membranes. They are key players in promoting cellular homeostasis, by maintaining membrane integrity, aiding in cell binding and adhesion, active and passive transport of substrates in and out of the cell, and many other vital functions. They come in many shapes and forms. One important superfamily of transmembrane proteins is the G-protein coupled receptors (GPCRs). The basic anatomy of these proteins is outlined in Figure \ref{hrh3_all}a. There are three main domains: the extracellular domain, the transmembrane (TM) region, and the intracelluar domain (ICD). The intracellular and extracellular domains interact with molecules in and out of the cell, respectively. As such, flexible loop regions tend to be found here, in addition to some segments of ordered secondary structures. The transmembrane region on the other hand is almost entirely made up of a bundle (seven in this case) of $\alpha$-helices forming a channel through the lipid bilayer of the cell membrane, allowing substrates to enter and leave the cell. The structures in Figure \ref{hrh3_all} include the incomplete X-ray crystal structure (b) and three computational models (c-d, obtained from the SWISS-MODEL repository \footnote{https://swissmodel.expasy.org/repository/uniprot/Q9Y5N1.}) of the H3 human histamine receptor (H3HR). 
Its primary function is the binding and release of the neurotransmitter histamine. As such, it is vital for normal neurological function and its impairment is associated with a number neurodegenerative disorders, including Multiple Sclerosis (MS), Alzheimer's disease (AD) and Parkinson's disease (PD) \cite{Rocha2014}. The crystal structure (PDB: 7f61) is missing the entire intracellular domain, approximately 105 residues. Transmembrane proteins like H3HR and other GPCRs are generally difficult to determine experimentally for several reasons. Part of this is due to the fact that these proteins exist in different media - the intracellular and extracellular domains exist in aqueous hydrophillic media (cytosol and plasma) while the transmembrane region is trapped in a hydrophobic environment surrounded by lipids. Experimentally, it is very difficult to provide a solvent for proteins that can emulate both environments. Thus, this class of proteins relies heavily on some sort of homology modeling for researchers to further investigate them. The ability of the deep learning methods to accurately fill in the gaps here has been investigated before \cite{Lee2022, Hegeds2022}. Hegedus et al \cite{Hegeds2022} found that AlphaFold2 performed well, but most of their structures appeared to mainly involve the ordered, transmembrane alpha helical bundles. Similarly, Lee et al \cite{Lee2022} found that AlphaFold2 and RoseTTaFold performed well in predicting the TM regions, but there were significant discrepancies when modelling the surrounding loop regions. This is precisely what we observe in our analysis of the H3HR models. The AlphaFold model is colored in green, a SWISS-Model based on isoform 7f61.1.A (essentially modelled without a template for the ICD) colored in orange, and another SWISS-Model based on isoform 5dsg.2.A (using the M4 muscarinic receptor as a template for all domains) colored in purple. The TM regions align quite well in all 4 structures as seen in Figure \ref{hrh3_all}c, and in all cases the RMSDs for this region is under 5.0 \r{A} (Figure \ref{hrh3_all}d). The differences in the ICD structures were much more pronounced between the models. The largest RMSD is observed between AlphaFold and both SWISS-MODELS (averaging 26.5 \r{A}). The RMSD in the ICD between the SWISS-MODELS is still significant (13.8 \r{A}), although about half as much as AlphaFold2's, where there appears to be virtually no order in the structure. In reality, the ICDs of these proteins are neither fully ordered nor fully intrinsically disordered, but rather a combination of segments of both, much like what is observed in the case of the purple isoform in Figure \ref{hrh3_all}d. Another interesting fact worth pointing out is that, for reasons which remain unclear, the AlphaFold2 model \footnote{https://alphafold.ebi.ac.uk/entry/Q9Y5N1.} is noticeably larger. The model includes an additional 26 residues at the N-terminus and another 14 residues at the C-terminus. With this, the model has 445 residues total, while the FASTA sequence for the H3HR crystal structure (PDB:7f61) contains only 407 residues, including those missing from the ICD. 

The ability to accurately generate full length in silico models for transmembrane proteins is both challenging and crucial. They remain among the most experimentally difficult to determine and are accordingly the most expensive human proteins to solve via crystallographic methods (with an average cost of \$2.5M per protein, at a 10\% success rate \cite{stevens}). Yet, solving these structures holds tremendous therapeutic value. As a clear example, approximately 35\% of FDA approved drugs target GPCRs \cite{Sriram2018}. 

\subsection{Multiple sequence alignment availability and diversity}
% \textcolor{red}{Shehab: BOTH 3.1 AND 3.2 HAVE FIGURES BUT THIS SECTION DOESN'T. Hakan: There was a figure, but we decided to move it to supplemental materials section due to figure limitation for the journal.} 

Given a single amino acid sequence, an MSA is a set of amino acid sequences generated from the initially queried sequence with respect to how similar sequences are to the initial sequence. MSAs capture important evolutionary information that are widely used in PSP, protein classification and protein design.

Using MSA-based information is not a new approach in PSP. Earlier works \cite{Pf_earlier_covar, Pf_earlier_bayesian}laid out the foundations for how one can use covariance methods to identify the co-evolving pairs of residues, essentially investigating the substitution patterns correlating among sequences from different organisms. Later on, more inferential techniques from this statistical summary have been applied to study protein-protein interaction from MSAs that can be used for analyzing spatial proximity of residues in the structure prediction \cite{Pf_messagepassing, PSICOV_covar_PF}. The underlying idea in these analyses, at its core, is that the residues that have co-evolved tend to stay in close proximity after the 3D structure stabilizes and achieves its native state. In recent years, one usually applies a statistical analysis to extract co-evolutionary information about the target protein and uses this information in a deep learning architecture to predict the 3D conformation. Over the years, there have been many works that utilize some type of neural network based architecture to achieve this \cite{deepCNNfor_PF, DLirrespectivecoevol_PF, AF2_paper, three_track_NN_PF}. However, the biggest breakthrough happened with the introduction of AlphaFold2 \cite{AF2_paper} where a transformer block ``Evoformer" was introduced that translates the structure prediction problem to a graph inference problem. This architecture is capable of capturing long-range dependencies among residues in the generated MSAs and successfully derives co-evolutionary information that produces highly accurate structure predictions.

With these new and more sophisticated deep learning techniques such as self-attention mechanism emerging, using the co-evolutionary information from the raw MSAs have become widely viable for PSP. Despite all these advances, access to diverse and deep MSAs present an important computational bottleneck. In particular, generating MSAs at a large scale is computationally expensive. Pre-computed MSA databases that contain millions of sequences, such as the one used in AlphaFold2, are not publicly available. In fact, a recent work \cite{ahdritz2023openproteinset} created an open-source database OpenProteinSet that contains more than 16 million pre-computed MSAs to make it accessible to the larger research community. This is a diverse and large database that researchers can use to train models for machine learning based structure prediction tasks. Hence, it is evident that MSAs play a central role for modern deep learning models for PSP.

It is also important to mention that in parallel with the rising popularity of large language models, protein language models (PLMs) are becoming a highly significant tool for PSP \cite{PLM_1, PLM_2, fang2023helixfoldsingle}. Most notably, MSA Transformer \cite{MSA_transformer} is a protein language model with a data set of 26 million MSAs that uses row and column attention mechanisms in its transformer based architecture. These language models are unsupervised models that predict the conformation of the protein based on the patterns recognized in the generated MSAs.

The information learned through MSAs by the deep learning models can potentially go beyond the co-evolution information as evidenced in \cite{MSA_energy_landscape}. It is hypothesized that the deep learning models can potentially learn an approximate energy function through MSAs and use this as a starting point in the energy landscape in the search for a global minimum. As a result, the learned energy function informs the search locally, giving an advantage to the model in this search problem. The authors support this claim by showing this learned energy function can be used to rank the accuracy of some target protein structures. This experiment reveals that, even though one uses a deep learning model that is not directly relevant to energy landscape information, it is still instrumental to understand the physics of the folding process and an efficient traversal of the energy landscape is necessary to identify the lowest energy conformation.      

\subsection{A multidimensional perspective on computational hardness}\label{perspective_section}
Protein sequence length does not singularly dictate prediction accuracy \cite{stevens2022benchmarking} but rather, a combination of factors such as mutations, disorder, and MSA depth collectively influences the ruggedness of the energy landscape. This complexity necessitates a broader scope of information beyond sequence length for a quantum solution to protein conformation problems. The fitness landscape in protein sequences, a critical metric for mutation analysis, directly influences biomolecular evolution by determining the Colony-Forming Unit (CFU) and approximating free energy \cite{hayashi2006experimental}. As mutations occur, local peaks emerge within the fitness landscape, leading to changes in the energy landscape, as depicted in Figure ~\ref{tc10b_tc5b_energy}, introducing uncertainty regarding its structure and potentially enabling quantum advantage. In comparison to structured proteins, Intrinsically Disordered Proteins (IDPs) possess shallow, rugged energy landscapes housing multiple local minima, especially in hypervariable regions \cite{cornish2020intrinsically}. This stands in contrast to ordered regions like the guanosine triphophate (GTP) binding domains, which exhibit a deep energy minimum. IDPs' sensitivity to environmental changes, including post-translational modifications, can significantly alter their energy landscapes by manipulating local minima and barriers, thereby creating uncertainty and ruggedness conducive to leveraging quantum advantages for conformation predictions. Additionally, studies on kinase families have shown that unregularized mutational statistics inference struggles to converge for shallow MSAs, indicating an increasingly rugged and ill-conditioned energy landscape \cite{haldane2019influence}. The results consistently highlight that reduced MSA depths result in Potts models reflecting more rugged energy landscapes, affirming the potential for quantum advantage in solving protein prediction problems affected by such increased ruggedness and uncertainty.

In this section, we present our view on the hardness of a PSP task in terms of three main parameters: the protein size (sequence length), the number of point mutations and the MSA depth. We discuss a general framework for identifying proteins for a structure prediction task, in particular using a physics-based quantum algorithm, which have high potential for yielding competitive results compared to state-of-the-art deep learning, MSA-based methods. We expect that this will be beneficial for both quantum researchers who plan to benchmark their novel quantum algorithms against state-of-the-art results, and also for biomedical researchers who wish to explore quantum capabilities to enhance their structure prediction results. Our perspective relies on a careful analysis of the current landscape of the PSP research and the most recent Critical Assessment of Structure Prediction (CASP) outcomes, a biennial event where the latest PSP methods are showcased and their performance is benchmarked \cite{kryshtafovych2021critical}. Our perspective is meant to serve as a general guideline for maximizing the best possible outcome in the regime where the best deep-learning based methods are known to perform poorly. While we identify this subspace of the problem space based on a particular quantum algorithm from \cite{PFIvano} and the actual boundaries of the subspace may change depending on the method adopted, we believe that the general framework provided persists if one wants to use an \textit{ab initio} method for PSP.

\begin{figure}[h!]
    \centering
    \includegraphics[width=16.5cm]{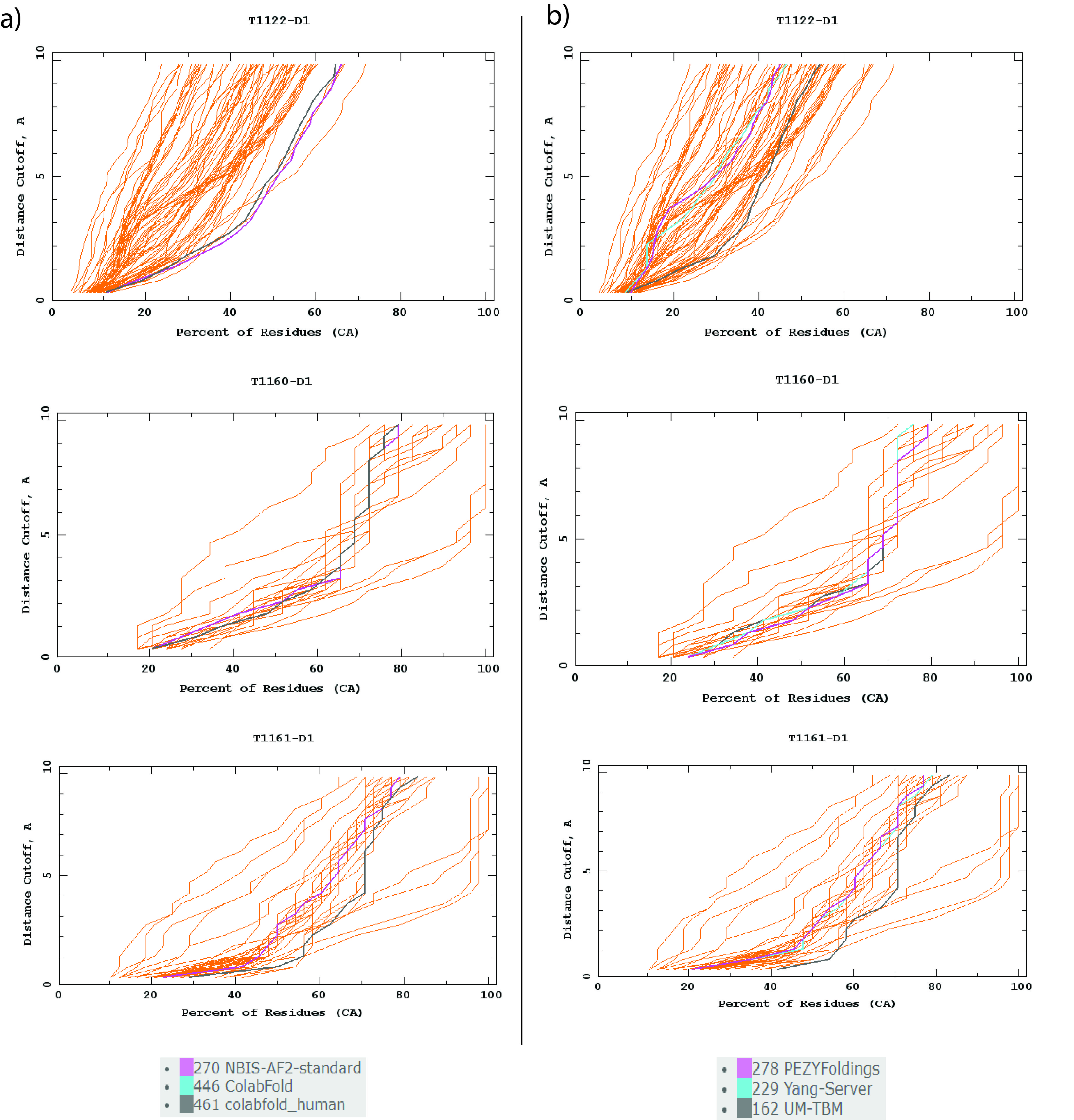}
    \caption{\textit{The plots highlight the performance of all groups for the target proteins, however we further distinguish; a) Performance of versions of AlphaFold2 that participated in CASP15 for the three targets. b) Performance of top three ranked groups with respect to overall z-scores in CASP15 for the three targets. The graphs show the percentage of residues in the target protein that are under a certain distance cut-off. It is easy to see that no group for the target protein T1122, no group was able to predict all residues under 10 angstroms. For T1160 and T1161, which exhibit point mutations, deep learning-based methods used by the groups listed has 80 percent of residues under 10 angstroms. In general, sharper increasing curves indicate poorer predictions. Images generated from CASP15 official website from the results.}}
    \label{t1122}
\end{figure}

In the previous sections, we discussed in detail three sources of hardness for the PSP problem and how they relate to energy landscape, in particular increasing complexity and ruggedness. Some of the target proteins from the most recent CASP15 provide further evidence for these hardness metrics. As discussed earlier, it is known that in the absence of deep and variable MSAs, DL methods fail to produce accurate predictions. In particular, it is reported in \cite{AF2_paper} that if the MSA depth is less than 30 sequences, AlphaFold2 prediction accuracy decreases significantly. Moreover, it is also mentioned that if the MSA depth exceeds 100 sequences, the algorithm has diminishing returns. Hence, one can conclude that for any protein sequence for which the generated MSA has depth less than 30, these deep learning based methods will yield low accuracy predictions. When measuring the performance of the predictions, there are two common metrics CASP utilizes; Global Distance Test - Total Score (GDT\_TS score) and Template Modeling score (TM score) \footnote{For detailed description, see https://predictioncenter.org/casp15/doc/help.html.}. Two target proteins T1122 and T1131 in CASP15 are examples of orphan proteins where a reliable MSA cannot be generated. In particular, none of the participants were able to produce a good prediction in the absence of MSAs. The best GDT\_TS score is less than 40 and and TM scores are around 0.5. A GDT\_TS score around 50 is interpreted to roughly predict the overall topology of the structure correctly, and similar interpretations for TM score as well. One can see for most predictions of T1122, only 60\% of residues are under 10 angstroms. The other target orphan protein T1131 has similar results for all participants with the best GDT\_TS score around 25 (this is usually deemed as good as a random guess) and the best TM score in 0.3s (See Figure \ref{t1122}).

Similarly, predicting the impact of the mutation is a difficult task. As detailed in \cite{CASP15_review}, detecting how the mutations alter the conformation was a big challenge in CASP15. In particular, two target proteins T1160 and T1161 differ by mutations at five residues. Both proteins were categorized as ``easy" since templates were provided and they are relatively short proteins with 48 amino acids. The top performing groups reached TM scores in 0.3 and 0.5s for the two target proteins respectively. Best GDT\_TS scores are 72.41 and 72.4 for the two proteins, while most groups achieved GDT\_TS scores in 50s. This is another example of how disruptive point mutations can be for a structure prediction task, in particular using a technique that requires reliable MSA-based information.

Based on these observations, if one wants to consider the parameters that can make a protein structure task difficulty, we propose that along with the protein size, MSA depth and the number of (point) mutations play a significant role. A detailed analysis is presented in the previous sections in terms of how they increase hardness and introduce more complexity in the energy landscape. In particular, our perspective is helpful if one chooses to use a method that does not rely on MSA-based information, such as physics-based or free modeling methods. In this work, we use the \textit{ab initio} quantum algorithm from \cite{PFIvano} to identify the subspace of the problem space where one can expect to produce predictions that are potentially more accurate than state-of-the-art, deep learning based methods.

The algorithm from \cite{PFIvano}, to our knowledge, is the best scaling quantum algorithm that constructs a 3D backbone of the protein from a single amino acid sequence. In this tetrahedral lattice model, a variational quantum algorithm is adopted to find the lowest energy conformation of the protein based on pre-calculated residue-residue interaction energies. We present a sequence of nested subspaces based on these three sources of hardness and the resource estimates for this particular quantum algorithm (See Section \ref{resource_est} for further details).

\begin{figure}[h!]
    \centering
    \includegraphics[width=16cm]{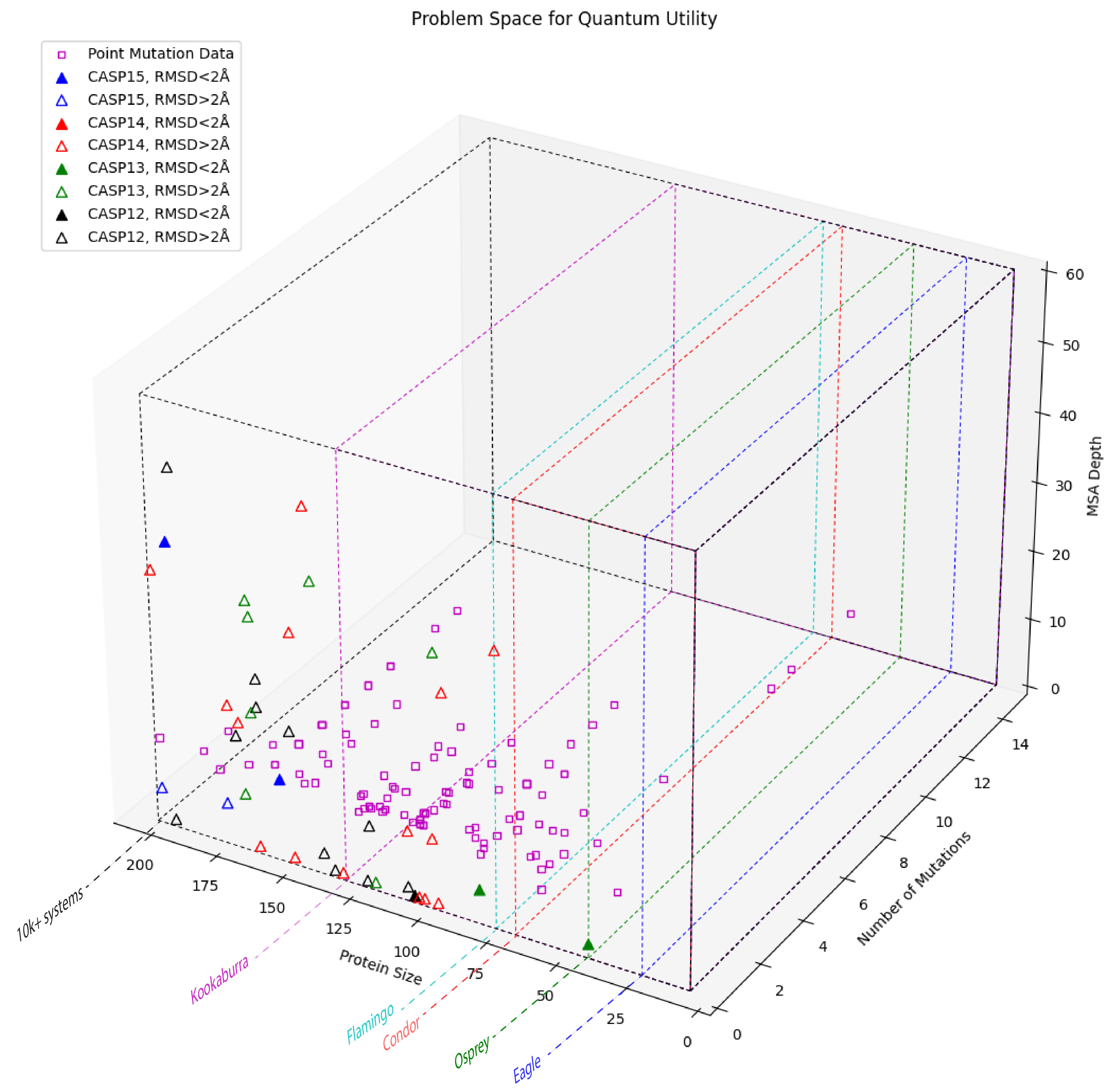}
    \caption{\textit{Subspace of PSP problems partitioned with different color rectangular boxes. This is in parallel with the IBM Quantum hardware road map as system size increases. The nested rectangular boxes represent the subset of proteins where deep learning based methods are known to perform poorly (data point markers with no facecolors), hence other \textit{ab initio} methods, including quantum algorithms, can potentially yield better predictions. CASP data is obtained from \cite{CASP_15_github_repo}, however sequence length data is added from CASP website directly. For each protein on the plot, average RMSD from top 10\% of the groups is calculated and added to the data. Point mutation data is obtained from \cite{thermomut}, and the MSA depth ($N_{eff}$) values are calculated using HHblits and HHpred tools \cite{hhblit_mpi}. Average RMSD values are not calculated for point mutation data set since they are not CASP targets. Clearly, there are many more proteins within these boxes. Our goal is to show that there are non-trivial, high-value targets for each of these regimes. The alignment with the IBM Quantum Roadmap is more of a symbolic representation. While we estimated qubit costs for the problem instances, one needs to perform a rigorous resource estimation for a concrete representation. This is beyond the scope of this work. We assume that for any range of qubit number, the quantum computer is able to perform a reasonable number of gates under reasonable timeframe.}}
    \label{hardness}
\end{figure}

Using the regular encoding of an amino acid sequence in \cite{PFIvano}, we observe that we can target proteins up to 22 amino acids in length using IBM's 127-qubit devices and extend this to 41 amino acids on IBM Osprey, a 433-qubit device. We have also estimated how this subspace gets larger as we have access to larger devices, aligned with IBM Quantum's hardware roadmap.\footnote{For further information see https://www.ibm.com/quantum/roadmap.} Since we have established that proteins with shallow MSAs are good candidates for this algorithm \cite{PFIvano}, we target the protein with MSA depth less than 60 as shown in Figure \ref{hardness}. MSA depth is measured by the number of effective sequences $N_{eff}$, similar to \cite{AF2_paper}, and the previous CASP data in the Figure \ref{hardness} is obtained from the public repository \cite{CASP_15_github_repo} and added some extra data curated by us. Since the target protein can exhibit any number of (point) mutations not exceeding the overall residue count, the protein size potentially determines the boundaries for the other axis representing the number of point mutations. However, the data we curated contains maximum 15 point mutations, so the axis boundaries are adjusted accordingly for better visualization. As a result, we argue that any target protein contained in the rectangular box regions in Figure \ref{hardness} is potentially a suitable candidate for a physics-based quantum algorithm to benchmark the performance and yield competitive results compared to state-of-the-art, deep learning template-based algorithms. 

We make a few important remarks and observations before concluding this section. Most importantly, while the actual numerical boundaries of the subsets provided in Figure \ref{hardness} depend on the quantum algorithm we adopt, the general axes that describe computational hardness and the underlying idea of ``going wide instead of going deep" provide a systematic way of benchmarking \textit{ab initio} quantum algorithms to maximize the outcome. As quantum hardware becomes more capable, we expect that this subspace with scale accordingly and include more potential target proteins. Moreover, with the current boundaries provided, this is a non-trivial subset of all proteins. There are 1321 protein structures in PDB that contains 22 or less residues. Similarly, almost 3000 protein sequences contain 41 or less amino acids in PDB database and if we query the condition that the sequences exhibit at least one mutation, we see that there are 509 protein sequences \footnote{All data collected from PDB website, note that some of these sequences can be part of a larger protein sequence. Since the proteins from PDB are already being utilized for training deep learning based algorithms, we note that these numerical values from PDB are meant to represent the variety of proteins that are available at these smaller scales.}.  

Benchmarking quantum algorithms and techniques is an important research area within the quantum community. Our perspective provides a novel and systematic approach that aims to maximize the utility of adopting a physics-based quantum algorithm and benchmarking the outcome. As a result, we believe that whether developing a quantum algorithm or exploring existing quantum algorithms for PSP tasks, selecting a target protein in the regions provided in Figure \ref{hardness} maximizes the utility and creates the best possible outcome.

\section{A quantum-classical hybrid workflow for protein structure prediction}\label{results-workflow}

While the promise of general quantum advantage comes with strong theoretical foundation, quantum computers are not expected to replace classical computers in every task. A scientific workflow may have a component that is computationally intensive and could be handled by a quantum algorithm while the remaining steps are treated on classical hardware. For example, simulating a catalytic reaction in the active site of an enzyme with a quantum Hamiltonian while the rest of the domains are modeled using classical algorithms (such as using conventional molecular mechanics force fields) \cite{Marx2021}. Initially, we envision quantum algorithms for PSP to be designed as part of hybrid quantum-classical workflows. Lattice models can help simplify the PSP problem by representing proteins as a chain of beads on a lattice. Generally speaking, these models employ an energy function to enumerate and score conformations, with a search algorithm used afterwards to identify the lowest energy conformation. The energy functions may range in sophistication, from a basic hydrophobic-polar model (where the alphabet is reduced to a binary representation, allowing for just three possible interaction types) to more detailed models considering various physical forces including the hydrophobic effect, electrostatic interactions, $\pi-\pi$ stacking, hydrogen bonding, induced dipole moments, and side-chain packing. In any case, the objective should always be to effectively reproduce the free energy associated with every possible type of interaction. The more detailed the energy function is and the more granular the lattice model, the greater the number of physically reasonable enumerated conformations. The challenge for the search algorithm then becomes the ability to effectively traverse this enormous space containing all possible conformations, which remains NP-hard even at a coarse grain resolution on a 2D lattice.

\begin{figure}[h!]
    \centering
    \includegraphics[width=16.5cm]{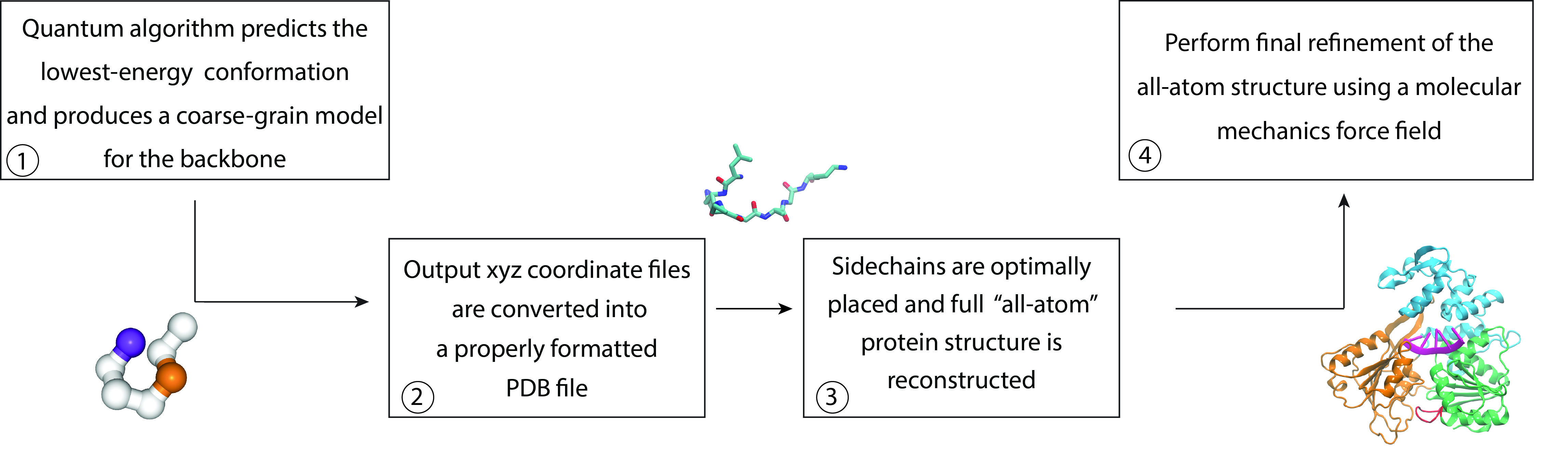}
    \caption{\textit{A schematic representation of the workflow used in this study. The most computationally demanding part, finding a coarse-grain representation of the lowest energy conformation of the protein structure, is performed on a quantum computer. The following steps are handled classically to convert the output into a desired format and post-process to construct the full structure, while preserving the quantum algorithm's originally predicted backbone geometry in the coarse grain model. A final refinement of the all-atom structure is then performed through further energy minimization using a molecular mechanics force field. This last step allows the protein to potentially reach an even more optimal configuration as the atoms and bonds are no longer constrained to the four turns of the original lattice structure.}}
    \label{workflow}
\end{figure}

The series and order of steps involved in this proposed workflow (see Figure \ref{workflow}, and \cref{supp-sec:workflow} of the supplementary information for a more detailed discussion of the workflow) are not uncommon in \textit{ab initio} PSP methods. For example, the Rosetta algorithm \cite{Leman2020} first creates small coarse grain fragments using a knowledge-based potential (the Rosetta energy function), much like what is done in the first step of this workflow (using the Mizayawa-Jernigan potentials instead \cite{MiyazawaJernigan}). These fragments are then assembled using a set of reference PDB structures until the full size of the sequence is built. In the later stages, the Rosetta ``all-atom'' energy function \cite{Alford2017} is then employed which refines the all-atom models. A clustering analysis is performed on the resulting models from each iteration traversing the search space, with the average structure from the largest cluster becoming the chosen solution. The fact that templates are used in this process is part of the reason why some members of the community have debated as to whether or not Rosetta is in fact a true \textit{ab initio} method, since it is mainly the fragments are constructed exclusively with an energy function. By now, we are of course aware that efficiency plummets when trying to do this beyond a few dozen amino acids. Nonetheless, incorporating the physics this way has clearly allowed the method to preserve accuracy while handling large sequences, an advantage over most other template-based methods. With the possible advantage quantum computing brings in handling large search spaces, fragment assembly may not be required, allowing for the enumeration and sampling of large conformers in a single pass of the algorithm. This is what our proposed workflow attempts to do - build the entire sequence in one step, sample the energies of all possible conformers, and finally refine the optimal solution. 

\begin{figure}[h!]
    \centering
    \includegraphics[width=12cm]{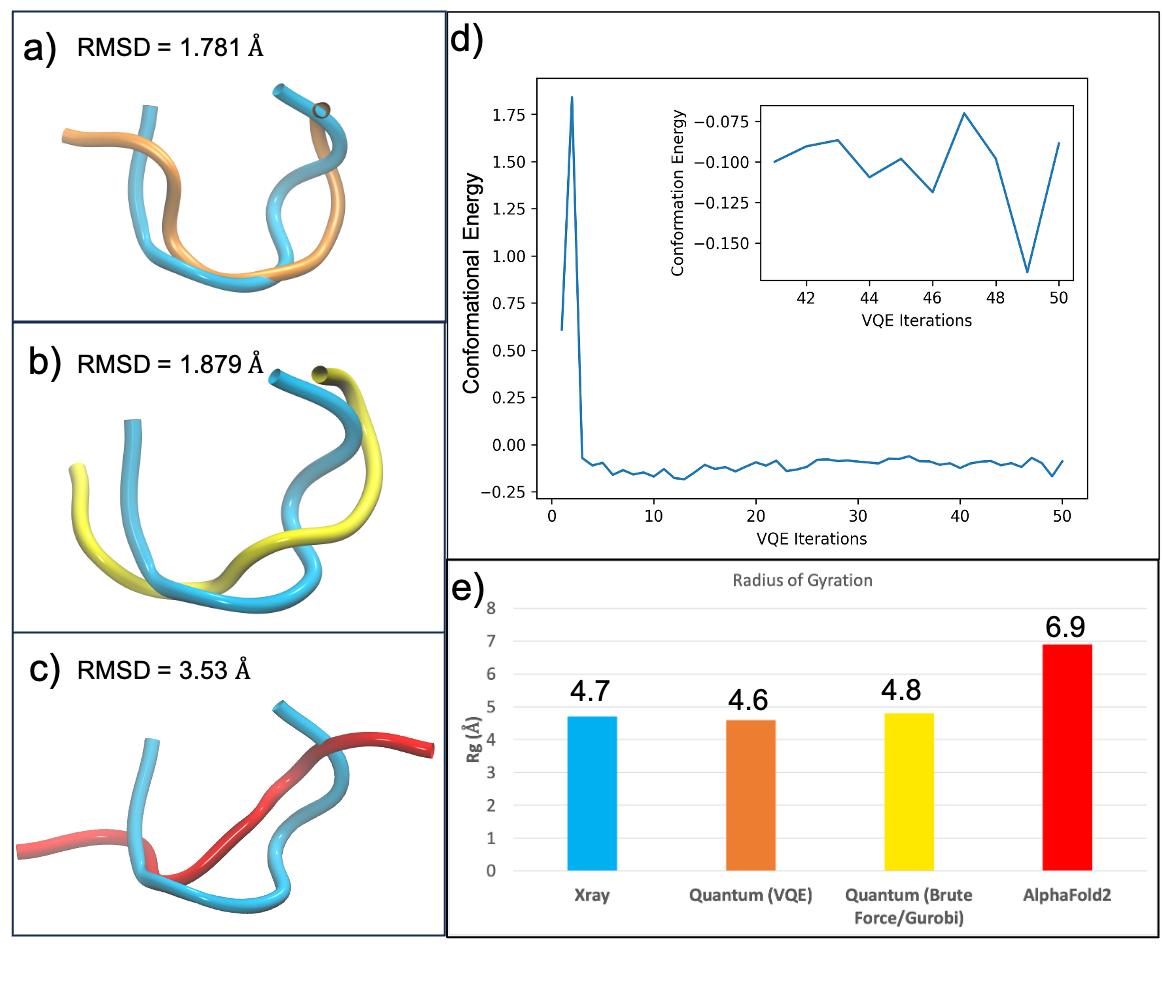}
    \caption{\textit{An initial validation of the workflow with the Zika virus helicase P-loop (LHPGAGK). In all cases, the coordinates from the experimental crystal structure are colored in cyan. The most accurate structure is produced by a) the quantum algorithm executed on IBM\_Cleveland, followed by b) the problem Hamiltonian solved classically by brute force and classical mixed-integer linear solver. The least accurate model was produced by c) AlphaFold2. This is observed by the relative RMSD values in each case, as well as e) the measured radius of gyration. The conformational energy plot in d) appears to demonstrate that VQE begins to continuously sample conformations around the basin after a handful of VQE iterations.}}
    \label{zika}
\end{figure}

We tested steps 1-3 of this workflow on a small, but highly relevant seven amino acid fragment (LHPGAGK) - the ``P-loop'' of the Zika virus helicase protein.  See \cref{supp-sec:zika_loop} of the Supplementary section for more discussion about the relevance of this protein in drug discovery. We chose the P-loop as the ``toy model'' for this workflow and present our results as an initial proof of concept. A comparison is made in Figure \ref{zika} between the results from a) the full quantum algorithm executed on IBM\_Cleveland, b) the quantum algorithm's Ising Hamiltonian (a classical cost function originally defined in \cite{PFIvano}) handled classically (by an exhaustive brute force search as well as Gurobi \cite{gurobi}), and c) AlphaFold2. In every case, the loop's experimental structure (extracted from the full crystal structure, PDB: 5gjb) are colored in cyan, while the models are colored in orange, yellow, and red. The backbones were aligned to the experimental structure and RMSDs were measured in each case using VMD \cite{Spivak2023}. The best model was in fact produced by the execution on the IBM\_Cleveland quantum device, with an RMSD of 1.781 \r{A} with respect to the crystal structure. During sampling, the conformational energy initially rises sharply but begins to continously hover around the basin shortly after a handful of VQE iterations (Figure \ref{zika}d). The Ising Hamiltonian produced by the quantum algorithm contains higher order polynomial terms (cubic and above), which can often increase the complexity of optimization in general, and more so on classical computers. We have also solved the same Hamiltonian classically to compare the performance. While the brute force search was performed on the native high order Hamiltonian, it  was also quadratized and converted into a QUBO (quadratic unoptimized binary optimization) Hamiltonian using a method developed by Fujitsu \cite{fujitsu}, and posteriorly solved with Gurobi. The Hamiltonian term corresponding to the second best result was obtained as the optimal solution by both Gurobi and a brute force search, yielding an RMSD of 1.879 \r{A} (Figure \ref{zika}b). AlphaFold2 produced the least accurate model (Figure \ref{zika}c), with an RMSD of 3.53 \r{A}, nearly twice as much as that predicted by the quantum algorithm. Radius of gyration was also measured in each case, and the same trend can be observed (Figure \ref{zika}e). The averaged results from IBM\_Cleveland (4.6 \r{A}) and the classical solvers (4.8 \r{A}) is the same as the experimental structure (4.7 \r{A}), while AlphaFold2's prediction is noticeably greater (6.9 \r{A}). 

These results highlight a few important things: 1) Both the tetrahedral lattice model and the Miyazawa-Jernigan potential employed in the algorithm \cite{PFIvano} could be sophisticated enough for applications in \textit{ab initio} PSP at this scale. 2) Even though VQE may not always find the absolute lower bound in the energy, it could be well suited for sampling near optimal low energy conformations that may even produce more accurate results than an exhaustive search when the measure of success is based on structural similarity to a ground truth experimental solution. It is well known that there could be several local minima that are quite close in magnitude and sign, yet be structurally distinct. An example of this can be seen in Figure \ref{tc10b_tc5b_energy}a, where the bottom of both funnels correspond to two distinct conformations similar in energy, but only one of them is closest to the experimental structure. 3) Classically rebuilding the all-atom structure can still preserve the originally predicted geometry by the quantum algorithm. 4) A quantum-classical hybrid workflow like this could possibly be used as a tool in tandem with template-based programs to predict binding domains characterized by loops. 

\section{Resource Estimation}\label{resource_est}

 Similar to determining which PSP problems may benefit from quantum computational advantages, accurately estimating the resources required for quantum algorithms that solve PSP problems presents its own difficulties. It is important to note that a rigorous resource estimation of a quantum algorithm for a given PSP problem is beyond the scope of a perspective paper. However, we contextualize our perspective with a simplified resource estimation framework. We only consider a utility-scale quantum algorithm however a similar framework can also be developed for a fault-tolerant quantum algorithm.

 For the discussion presented in this section, we operate under the following set of assumptions:

\begin{itemize}
    \item We use IBM quantum computers as our target devices. The diversity in QPU qubit connectivity and topology can have an impact on the resource estimation. Our estimates are based on the 127-qubit Eagle (R3) quantum chip on IBM\_Washington.
    \item The quantum algorithm employed changes the various resources needed to execute the experiment. We implement the algorithm from \cite{PFIvano}, which is a variational quantum algorithm. As a result, qubit count depends on the choice of lattice for the coarse-grain model and the interaction model between residues. Moreover, the ansatz we have used is the RealAmplitudes ansatz from Qiskit's circuit library for only one repetition. We point the readers to \cite{du2022efficient} for comparative overview of different types of ansatze. 
    \item A more comprehensive resource estimation requires analyzing the entire workflow in Figure \ref{workflow}, whereas we only focus on the first step of the workflow. Furthermore, while we analyze some commonly used metrics such as qubit count, circuit depth, ECR depth, and number of measurements, there are other parameters one can consider such as overall and gate-level execution time, overhead from error mitigation, etc.
    \item In a future work, our goal is to provide a robust resource estimation for PSP, where we consider the other parameters mentioned above.
\end{itemize}

\begin{figure}[h!]
    \centering
    \includegraphics[width=16.5cm]{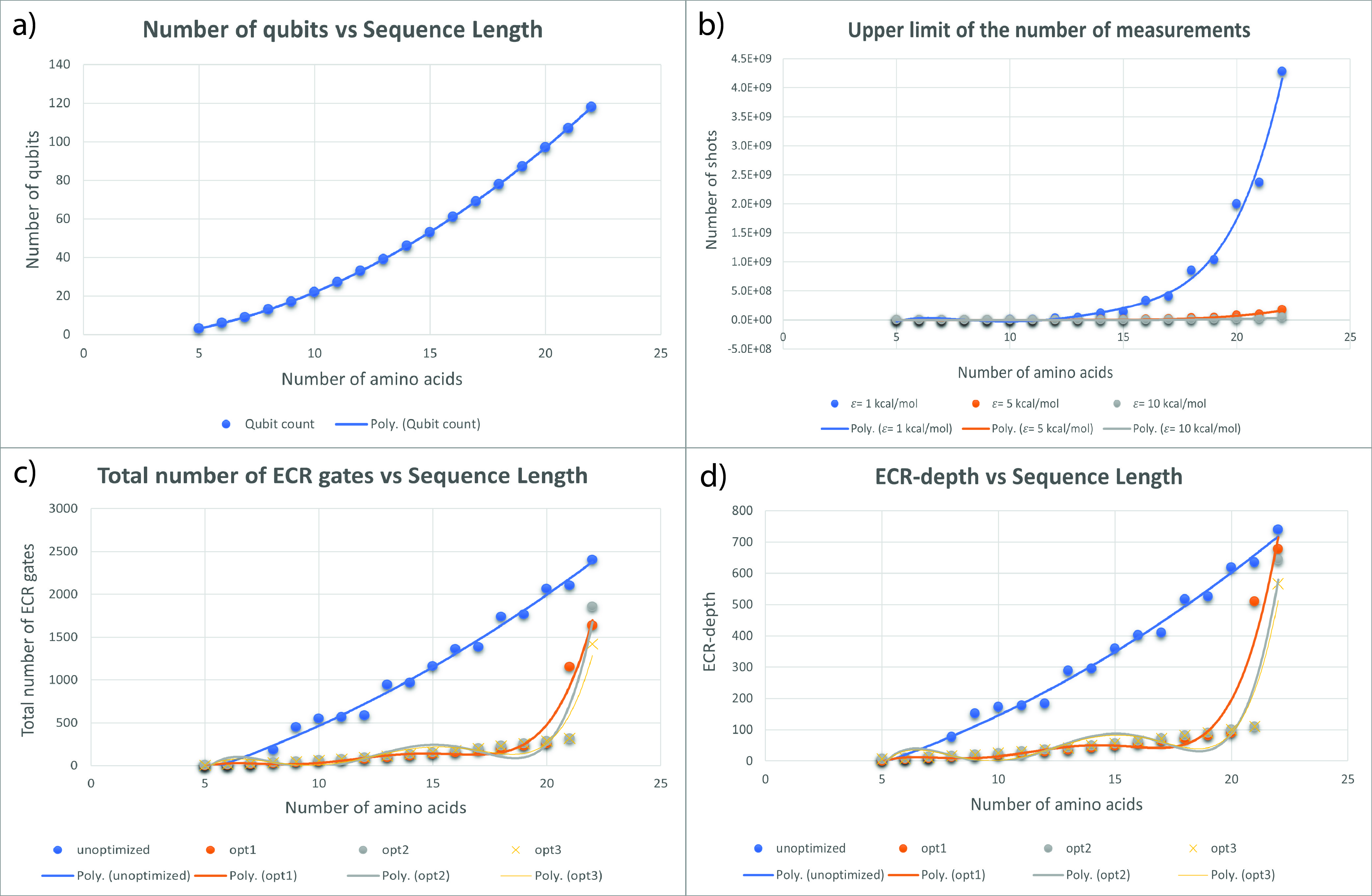}
    \caption{\textit{a) The total number of qubits scale quadratically as the protein size increases. We consider both configuration and interaction qubits to encode a given amino acid sequence, and analyze the scaling. b) We estimate an upper bound for the number of measurements needed to predict a protein structure within a fixed energy error margin using the work from \cite{vqe_shotnumber}. The energy unit is converted to kcal/mol from Hartree for consistency with our workflow. See Section ~\ref{supp-sec:measurement-formula} of the supplementary materials for mathematical formula to calculate the upper bound. The three plots show different upper limits to predict a protein structure within $\varepsilon=1,5,10$ kcal/mol range of the lowest energy conformation. The upper bound is not known to be tight. We further expect the empirically sufficient number of measurements to be significantly lower than this upper bound. c) Shows the total number of ECR gates in the circuit as the protein size increases for different optimization levels. Qiskit transpiler allows four levels of optimization which are defined in \cite{ibm_doc}.  d) Shows the ECR-depth as a function of protein size.}}
    \label{resource}
\end{figure}

For simplicity, we have adopted a simple regression model for extrapolation. Our approach involves mapping the problem to a quantum circuit, characterizing the circuit in terms of the number of gates and qubits required, applying different levels of circuit optimization and analyzing the number of ECR gates required for execution. We derived our estimates using the sequence of an experimentally verified protein structure (membrane-proximal cytoplasmic domain tail of platelet integrin $\alpha$IIb-$\beta$3) \footnote{The PDB entry can be found here https://www.rcsb.org/structure/1M8O.})
The estimates provided in the Figure ~\ref{resource} were calculated  for up to 22 amino acids. Using the regression model we built, we have established the resource cost in different categories such as number of qubits, number of ECR gates that can be executed in parallel, total number of gates, total circuit depth for both optimized and unoptimized circuits, and upper limit on the number of measurements.

Our estimates show that there is a perfect quadratic relationship between the protein sequence length and the number of qubits required. The primary reason we have based our regression model up to 22 amino acids is the current capabilities of the quantum hardware. Our estimates show that for a protein sequence with 22 amino acids, the number of qubits needed is 118 (see Figure \ref{resource}a). Extrapolated according to these estimates, we expect that we can go up to 41 amino acids with the release of IBM's Osprey, 433 qubit quantum computer, and up to 67 amino acids on IBM's Condor, a 1121 qubit machine. As a comparison, similar estimates were conducted when converting the Hamiltonian into a QUBO format and this can be found in \cref{supp-sec:hamiltonian} of the supplementary materials. While qubit count is an important constraint for hardware experiments, we are also assuming that we will be able to run the number of gates that we project in these estimates. As we have access to more qubits and higher gate fidelity, we believe that it will be possible to predict larger and larger protein structures using quantum algorithms. While in terms of protein size, current deep learning based methods can target much larger structures, as we discussed in Section ~\ref{main_section}, we believe that there are still regimes in smaller proteins where physics-based quantum algorithms can yield better results. 

In conclusion, we have established an initial analysis of the quantum computational resource cost for protein conformation predictions using the quantum algorithm from \cite{PFIvano}, based on the full protein structure of platelet integrin $\alpha$IIb-$\beta$3 cytoplasmic domain \cite{Vinogradova2002}. Our estimates are crucial for understanding the limits of the subset of the problem space we want to target. There are various other techniques such as dense encoding, optimizing logical-to-physical qubit mapping, tensor-based circuit transpilation methods, circuit cutting and dynamic circuits that can enable us to predict larger proteins. In our future work, our goal is to explore some of these methods to enhance the performance of the quantum algorithm.

\section{General discussion}
The protein folding problem and protein structure prediction remain challenging tasks despite advances. Template-based methods, particularly those propelled by machine learning, have enabled researchers to obtain realistic models faster than wet lab experiments, establishing structure-function relationships that would have otherwise taken a very long time to discover. Growth of databases like the Protein Data Bank, now exceeding 200,000 experimental structures, underpins the success of these methods. However, known sequences outpace solved structures, around 300 million versus 200,000 \cite{Jisna2021}. This highlights expansive unseen biodiversity. Advancing physics-based PSP methods is crucial to help bridge this gap.

Some argue that most natural folds have already been discovered, and by extension the template-based methods should be capable of predicting most structures. Remarkable performances like those of AlphaFold2 support this claim, given enough homology. However, improvement is still needed when dealing with mutations (even in conserved regions), orphan proteins, shallow MSAs, and novel sequences. AlphaMissense may address mutations \cite{alphamissense}. Still, there are many missing pieces in this puzzle according to many in the biophysics community. Overcoming these limitations fundamentally requires simulating physics. Molecular dynamics could be a solution, but is constrained by timescales, system sizes, and the chosen force fields. \textit{Ab initio} methods also scale poorly on classical hardware as the search space grows unmanageably large. 

Quantum computing shows promise in both accuracy and scalability, as seen by the results of our quantum-classical workflow in predicting the Zika virus NS3 helicase P-loop. It produced an accurate model with a backbone RMSD of less than 2.0 \r{A} compared to the experimental structure, and with superior scalability versus classical optimization \cite{PFIvano}. With 10,000 qubits and sufficient circuit fidelity, prediction of biomedically relevant proteins and mutants could be attainable. For instance, our estimates indicate hemoglobin's 141 residues may be predictable using 4,967 entangled qubits – an extremely valuable prospect since its structure enables fundamental life functions. Although this structure has long been determined experimentally, it is at this size and scale where some of the most biologically relevant proteins exist.

Early quantum algorithms will likely mirror classical physics methods. Eventually quantum chemical calculations could replace knowledge-based energy functions, initially with DFT or Hartree-Fock for larger peptides than what is currently possible on classical hardware. The ultimate goal is perhaps simulating exact electron dynamics, deriving molecular and electronic structures in real time, capturing nature in its most absolute form. This could bring about a new era for molecular force fields as well, perhaps yielding the ability to not only predict optimal protein structures irrespective of path, but truly help solve the path-dependent protein \textit{folding} problem itself. 

The future of structural biology depends profoundly on developing and scaling up efficient physics-based PSP methods. Quantum computing’s potential to simulate nature's fundamental mechanics may help overcome the historical barriers. This will not have to be mutually exclusive, but could be rather synergistic, with quantum PSP methods helping complement the strengths of classical template-based methods, and vice versa. Multidisciplinary collaboration between biophysics, chemistry, computer science, and structural biology can help unravel the remaining mysteries. For the first time in decades, elucidating how and why proteins fold appears within reach, presenting a milestone deeply advancing our comprehension of the subtle intricacies enabling life.

\section*{Acknowledgments}
Portions of this material is based upon work supported by the Defense Advanced Research Projects Agency (DARPA) under Contract No. HR001122C0102 to O.S. Any opinions, findings and conclusions or recommendations expressed in this material are those of the author(s) and do not necessarily reflect the views of the Defense Advanced Research Projects Agency. The authors thank  Kenneth Merz, Iris Smith, and William Martin from the Cleveland Clinic, and Emily Pritchett, Saif Rayyan, Antonio Mezzacapo, Courtney Long, Wiktor Mazin, Niall Robertson, Michal Rosen-Zvi, Jeannette Garcia, Filippo Utro, Kahn Rhrissorrakrai, Ryan Sweke, Joseph Morrone, Yoel Shoshan, Ivano Tavernelli, Sarah Mostame, Sara Capponi, Jannis Born and Cynthia  McKinney from IBM, and Robert Jernigan from Iowa State University
of Science and Technology for their expert opinions and insightful feedback on the manuscript or during the project. They also acknowledge DARPA, NASA, and the MIT Lincoln Lab team involved in the Phase 1 of the DARPA Quantum Benchmarking program led by Joseph Altepeter, for rigorous feedback.

\section*{Competing interests}
The authors declare no competing interests.

\bibliographystyle{unsrt}
\bibliography{bibliography}

\end{document}

\title{Supplementary Information for ``A perspective on protein structure prediction using quantum computers"}

\author[1]{Hakan Doga}
\author[2]{Bryan Raubenolt}
\author[2]{Fabio Cumbo}
\author[2]{Jayadev Joshi}
\author[2]{Frank P. DiFilippo}
\author[2]{Jun Qin}
\author[2]{Daniel Blankenberg}
\author[3]{Omar Shehab}

\affil[1]{IBM Quantum, Almaden Research Center, San Jose, California 95120, USA}
\affil[2]{Center for Computational Life Sciences, Lerner Research Institute, The Cleveland Clinic, Cleveland, Ohio 44106, USA}
\affil[3]{IBM Quantum, IBM Thomas J Watson Research Center, Yorktown Heights, NY 10598, USA}
\date{}

\maketitle

\section{A brief introduction to Protein Structure Prediction}\label{sec:PSP_intro}

While the structure of DNA and RNA remains largely helical irrespective of the sequence, proteins on the other hand may form a nearly limitless diversity of tertiary structures. This is due in part to the fact that nucleic acid sequences are limited to just 4 fundamental building blocks (adenine (A), cytosine (C), guanine (G), and uracil (U) or thymine (T) for RNA or DNA, respectively). The translated proteins from these nucleic acids are made up of a much more diverse set of building blocks; a set of 20 amino acids varying in charge, hydrophobicity, polarity, and size. Because of this diversity in the chemical space, the possible structures of proteins go far beyond the hydrogen-bond stabilized helical moeities adopted by nucleic acids. Thus, even for the same number of monomers, the number of possible conformations a protein could adopt is orders of magnitude higher than that of a nucleic acid. Of course, where the problem becomes perpetually harder is the fact that proteins can be thousands of amino acids in length. In this section, we discuss some of the best current classical structure prediction methods and their limitations, as well as some of the latest approaches in quantum computing. In general, these classical methods fall into 1 of 2 categories: 1) either being template-based (meaning that the method generates models by somehow using information from templates in a database of experimentally determined structures) or 2) the physics-based methods (often regarded as ``free modeling`` approaches). 
\subsection{Physics-based or Free Modeling Methods}
Molecular Dynamics (MD) simulations provide atomic trajectories of the dynamic time-evolution of chemical systems (including proteins in their physiological environments) by applying Newton's laws of motion and a force field (a set of radially dependent parameters describing the potential energies associated with the interactions between unique atom types) to a set of initial atomic coordinates \cite{MD_sim}. As such, these simulations can be used to not only find an optimal conformation, but also shed light on the folding process itself. This can be done by starting the simulations at a higher temperature in an unfolded state, and gradually lowering it to physiological conditions. These simulations can also be performed near the theoretical melting point (starting from folded or unfolded configurations), where the proteins can be observed to unfold and refold over time. These methods can provide rather realistic models, with several studies reporting RMSDs of less than 2.0 angstroms when compared to experimental structures \cite{anton4}. However, the complexity of these simulations, due to factors such as protein size and water model (and number of water molecules), requires intensive computation. This makes MD impractical when initial conformations are far from their energy minimum. Simulations from unfolded states to a presumed low-energy conformation are typically limited to small peptides (less than 80 amino acids), hindering scalability for most biomedically relevant proteins \cite{Pang2017, Piana2013, LindorffLarsen2011}. Furthermore, these studies often necessitate high-end compute resources and lengthy time scales, making them impractical for many researchers.  

\textit{Ab initio} protein structure prediction methods, including \textit{de novo} design methods, predict protein structures from amino acid sequences by enumerating conformers and exploring a vast conformational space. Often using knownledge-based or statistical potentials to describe the energies of each conformer, a search algorithm is usually employed to find the structure with the lowest free energy, using methods like simulated annealing, genetic algorithms, or Monte Carlo methods \cite{de-novo}. Although sometimes regarded as a hybrid approach and not a true \textit{ab initio} method, the Rosetta software is a renowned tool which uses knowledge-based potentials in its two energy functions (coarse grain and all atom energy) \footnote{Details can be found here https://github.com/RosettaCommons/.}. However, similar to MD simulations, these methods succeed mostly with small proteins as computational demand grows with the number of amino acids. An instance is the prediction of the 112 residue T0283 protein, which, while accurate, required significant time and computing power \cite{Qian2007}. It was computed using Rosetta@Home, a distributed resource version of Rosetta, on about 70,000 home computers over nearly 2 years. It is virtually impossible to perform exhaustive searches of the conformational space for reasonably sized proteins, which is why most of these methods end up using heuristic methods. This underscores the inefficiency of classical computers for physics-based prediction algorithms.

\subsection{Template-based methods}
 The limitations discussed above and the growth of experimental databases like the Protein Data Bank (PDB) led to the birth and evolution of template-based methods. These include threading methods (for example ITASSER\cite{itasser}), homology modeling programs (such as SWISS-Model\cite{swissmodel}), and the now well known Deep learning (DL) methods (such as AlphaFold2\cite{Jumper2021} and RoseTTaFold\cite{three_track_NN_PF}). DL is another powerful tool for protein structure prediction and has been increasingly used in recent years. These algorithms can learn patterns from large data sets, and they have been used to predict aspects of protein structure, such as secondary structure, solvent accessibility, contact maps, leading to impressively accurate full 3D structures. AlphaFold2 uses deep learning techniques with a transformer-based architecture to predict a protein’s structure from its amino acid sequence. It was trained on a vast data set of known protein structures (which in itself is quite expensive to do), nearly the entire Protein Data Bank (PDB), and uses this knowledge to predict the structures of unsolved proteins.

\subsection{Protein conformation prediction with quantum computing}
Each of the previously discussed methods has strengths and weaknesses. MD simulations and \textit{ab initio} physics-based methods can potentially provide very accurate results but are usually computationally intractable at scale. DL methods can be much faster and are showing increasingly good accuracy, but they depend on having a large data set of known structures to learn from, and they may struggle with proteins that are very different from those in the training data set. The need for better computational methods is evident from the fact that while there are approximately 12,000 experimentally verified protein structures uploaded to PDB every year, with a little over 200,000 structures in total, there are over 250 million known protein structures based on sequencing data \footnote{Data obtained from https://www.uniprot.org.}, with this gap continuously growing larger every year. Thus, innovative methods such as quantum computing need to be explored.

There have been numerous works in the last couple of decades that utilize quantum computing to find the lowest energy conformation of a protein structure, building simplified lattice models to represent the predicted structure. The early work of Perdomo-Ortiz et al. \cite{perdomo1, perdomo2} defined an Ising-like Hamiltonian for adiabatic quantum computing to construct 2D lattice structures using the hydrophobic-polar (HP) model of proteins, and implemented the algorithm on a quantum annealer for tetrapeptide and hexapeptide chains. Later works \cite{babbush, babej2018coarsegrained} have introduced certain improvements in terms of quantum resources needed on coarse-grained 3D lattice models, however remained unattainable for predicting larger peptides on quantum annealers. Using a classically predicted backbone structure, \cite{Mulligan752485} proposed using a quantum annealer to identify side chains and various conformations, particularly for designing proteins. A more efficient quantum algorithm proposed by Robert et al. \cite{PFIvano} defines a model Hamiltonian for predicting structures on a coarse-grained tetrahedral lattice model, and the authors demonstrate the algorithm on two peptides, 7 and 10 amino acids, using a variational quantum algorithm. A more recent work \cite{nathanwiebe_pf} proposed a pure quantum method that leverages Grover's search algorithm with an adder-only oracle to find the lowest energy conformation of a given protein structure. The table below shows a comparative picture for lattice-based quantum algorithms for protein folding problem \ref{compare}.

\begin{table}[h!]
    \centering
    \small
\begin{tblr}{ ||Q[c,m]||Q[c,m]||Q[c,m]||Q[c,m]||Q[c,m]||Q[c,m]||Q[c,m]|| }
 \hline
 Model & HP\cite{perdomo1} & CG\cite{perdomo2} & CG\cite{babbush} & CG\cite{babej2018coarsegrained} & CG\cite{PFIvano} & HP\cite{nathanwiebe_pf} \\
 \hline
 Lattice  & all & all & all & all & Tetrahedral & 2D  \\ 
 \hline
 Types  & 2 & $\infty$ & $\infty$ & $\infty$ & $\infty$ & 2 \\ 
 \hline
 Interactions & Nearest & Nearest & Nearest & Nearest & $l^th$ Nearest & Nearest \\ 
 \hline
 Locality  & log$_2N$ & $N$ & 4 & $N$ & $l+2$ & log$_2N$ \\
 \hline
 Qubits & $N$log($N$) & $N^2$log($N$) & $N^3$log($N$) & $N$ & $N^2exp(l)$ & $N$log($N$)  \\ 
 \hline
Scaling & $N^8$ & $exp(N)$ & $N^{12}log^4(N)$ & $exp(N)$ & $N^4$ & $\sqrt{N}$ \\ 
 \hline
\end{tblr}
\caption{\small\textit{The comparison of different work on protein folding using quantum computing. HP refers to Hydrophobic-Polar and CG refers to Coarse-Grain. The table recreated from \cite{PFIvano} and the values of the last column added for \cite{nathanwiebe_pf} based on IBM-SP model presented.}}
    \label{compare}
\end{table}

\section{The Levinthal paradox and the exponential growth of the search space}\label{sec:levinthal}
Proteins come in many different shapes and sizes, from small designer peptides less than 10 amino acids in length, to the largest human protein, Titin, which is comprised of over 37,000 residues \cite{titin}. There is an inherent dependence between producing an accurate structure, and the length of the sequence. In NMR experiments for example, the larger the structure, the more clustered the amino acids could be, which could often lead to resonance degeneracy, making it more difficult to discriminate between atom types. Computationally, it becomes a giant puzzle with many more pieces needed to solve. As mentioned previously, the size of the conformational space (the number of shapes the protein could adopt) dramatically increases with size. In a way, increasing the sequence length is sort of like increasing the dimensions of a Rubik's cube - more rows of blocks lead to many more possible configurations to sample, increasing the complexity of the problem and the time to arrive at a solution.

 The overall tertiary structure of a protein is influenced by interactions rising from both the backbone as well as the sidechain groups. The backbone is made up of the three rotatable bonds (and their respective angles, See Figure 2b of the main text) that define a peptide: 1) the $\phi$ angle between the backbone amine group and the $\alpha$-carbon, 2) the $\psi$ angle between the $\alpha$-carbon and the carbonyl group, and the $\omega$ angle at the center of the peptide bond itself, between the amine and carbonyl groups. The $\omega$ angle generally presents little deviation and is kept in a stable \textit{trans} configuration (180 degrees). The $\psi$ and $\phi$ angles on the other hand, are more flexible and largely influence the backbone's modularity. In general, a protein with \textit{n} amino acids will have \textit{n}-1 peptide bonds. With each peptide bond having rotatable $\psi$ and $\phi$ angles, this means that there are two degrees of freedom to consider per peptide bond. This translates to a total of 2(\textit{n}-1) degrees of freedom for a protein of \textit{n} amino acids. Even if one were to simplify the problem by limiting the number of variations in these angles to produce no more than three distinct conformations per peptide bond, the size of the conformational space dramatically increases even for the smallest of sequence lengths (total possible conformations under these constraints scales as 3$^{2(n-1)}$). This simplification of course also ignores the conformational variability in the sidechains, which often contain several rotatables bonds as well. The sidechain flexibility gives way to a series of interactions with the solvent, ions, and neighboring sidechains, which ultimately impacts the formation of the overall secondary and tertiary structures. To put things into perspective though, by only considering the backbone's two degrees of freedom, the three possible conformers per peptide bond, and only spending 1 picosecond sampling each conformer, for a 100 amino acid protein (which isn't too big for all practical purposes), an exhaustive search to find the most optimal (i.e. the most thermodynamically stable) conformation among the total number of possible conformations, exceeds the age of the universe in years. This is famously known as Levinthal's paradox \cite{levinthal}, and this relationship between sampling time and protein size is plotted in Figure 2a of the main text. This is paradoxical because in reality, most proteins fold to their native states in the order of microseconds to milliseconds. While it has been suggested that computationally biasing towards the native states may reduce this time significantly \cite{levinthal_karplus, Martnez_levinthal}, it remains an intrinsically NP-hard problem computationally \cite{pf_nphard}. 

Levinthal's paradox tells us that classical exhaustive searches are not possible. On the other hand, heuristic search algorithms and optimizers could offer good solutions without guarantees of finding the optimal one. With quantum computing, the idea would be that several possible conformations can be represented by several quantum states in superposition, allowing for a more ``simultaneous" sampling of the space, so to speak. This could certainly present advantage in tackling the search problem aspect of PSP. Yet, finding this optimal solution is only as good as the thermodynamic description of the problem itself. The solution could be optimal among the rest, but it may still be physically inaccurate. The true nature of protein folding is arguably a quantum chemical process (at least with respect to the behavior of electrons, which are at the center of hydrogen bonds, the Van der Waals interactions and induced dipole moments). To truly simulate this phenomenon, a quantum hamiltonian accounting for all correlated electrons among interacting residues may be the key in the future. Due to quantum resource limitations in the NISQ era, this is still not achievable. For all practical purposes, however, a classical thermodynamic hamiltonian (such as that found in \cite{PFIvano}, using statistical contact potentials from \cite{MiyazawaJernigan}) could be sufficient to approximate the interactions and predict the resulting folds. Perhaps using quantum computers to simulate it and efficiently sample the phase space will provide the necessary speedup in the near term. 

\section{The ``P-loop'' of the Zika virus NS3 helicase protein}\label{sec:zika_loop}
The helicase is a motor protein tasked with unwinding the viral double stranded RNA into single strands, a vital step in the replication process and infection life cycle of the Zika virus \cite{Raubenolt2021}. This happens through ATP hydrolysis, whereby the chemical energy released during this process is converted to mechanical energy, inducing a series of conformational changes in the RNA-binding domain that in turn cause the unwinding of the nucleic acid. The P-loop plays a key role in coordinating the ATP molecule and catalyzing this reaction. It is a small 7-10 amino acid loop that is highly conserved in many RNA helicases of the same family, including those found in coronaviruses \cite{Raubenolt2022}. Binding of a ligand at this site results in competitive inhibition against ATP, which renders the protein inert. The same can also be achieved through allosteric modulation by a ligand, where the induced conformational changes at the P-loop lead to a loss of catalytic activity. Thus, it is no surprise that viral helicases are being investigated as promising targets of antiviral therapies \cite{Frick2003, Spratt2021, Halma2022}, both those targeting the primary active site (P-loop and neighboring residues), as well as allosteric sites. For intelligent drug design of helicase inhibitors, an accurate representation of the chemical space around the P-loop, RNA-binding domain, and any releveant allosteric residues, must be obtained. 

\section{Structures for quantum speedup}
\label{sec:quantum}
There is a ``hierarchy of structure" when it comes to the potential for exponential quantum speedups in the black-box model \cite{aaronson2022much}. At the bottom are problems with no imposed structure, where only polynomial speedups are possible which is also known as the Grover speedup \cite{grover1996fast}. The canonical problem in this category is finding a marked item in an unstructured database. Optimization over a rugged energy landscape is another example. The immediate next class of problems only have minimal structure such as the hypothetical problem of a black-box being either a 1-to-1 or 2-to-1 function. The task is to determine which one. Known as the {\it collision problem} in theoretical computer science, its applications include finding collisions in cryptographic hash functions, scattering data evenly in a hash table, detecting MAC address collisions on a network, pseudo-random number generation, URL shortening, plagiarism detection, etc \cite{aaronson2004limits}. The second class of problems from the top has the structure known as non-Abelian hidden subgroup structure. For a given black-box function, the task is to query the black-box using polynomial amount of resources to determine whether there is a hidden subgroup inside the non-Abelian group being act on by the black-box function. Examples of such problems include graph isomorphism \cite{ettinger2004quantum, grigni2001quantum, shehab2017quantum}.   At the top are problems with Abelian group structure  \cite{lomont2004hidden}, which typically allow exponential speedups through algorithms like Shor's \cite{shor1994algorithms}. It should also be noted that quantum algorithm research is an active field, and it is not unusual to expect more information to emerge over time impacting the hierarchy \cite{lomonaco2007grover}. 

NP-hard problems are characterized by an exponential growth in difficulty as the problem size increases \cite{np-hardness}. Protein conformation prediction, which is known as an NP-hard problem \cite{unger1993finding}, maybe considered both as an optimization problem or a search problem. Hence, the worst-case scenario can be reformulated as a search in an unstructured database where the globally minimum energy is a marked element. Therefore, the upper limit of quantum speedup is quadratic and the problem lies in the bottom layer of the above-mentioned hierarchy. In a quantum search algorithm, there are typically four stages - initialization, database loading, probability amplification through interference and search result readout. In quantum algorithm jargon, they are also known as ``initialization, Grover’s oracle, Grover’s diffuser and measurement" respectively \cite{khatami2023gate}. In a classical search algorithm, one has to explore all search paths in the worst case. Using a quantum algorithm may be counter intuitive since encoding a database of discrete or real entities into the exponentially larger Hilbert space should've made the haystack bigger to find the needle in.  However, quantum mechanics allows us to associate negative probabilities to a subset of these search paths which eventually reduces the problem space and provides Grover speedup which is quadratic.  There is no analog for this in classical computer science.

\section{The quantum algorithm}\label{sec:workflow}
\subsection{Details of the workflow}\label{sec:workflow}

The quantum algorithm we've employed from \cite{PFIvano} is open source and currently part of Qiskit Research. It uses a tetrahedral lattice model to produce a coarse grain representation of the protein, a statistical potential for establishing the interaction energies between residue contact pairs \cite{MiyazawaJernigan} and a variational quantum eigensolver (VQE) to search for the resulting lowest energy conformation in the configuration space. The model can represent branched heteropolymers consisting of N monomers on a tetrahedral (``diamond") lattice. This choice of lattice is justified by its alignment with the chemically plausible bond angles (109.5\degree) and dihedral angles (60\degree and 180\degree) naturally found in proteins. There are three main functions to the Hamiltonian: one tasked with handling growth constraints $H_{gc}$, another enforcing chirality $H_{ch}$, and a final energy function reproducing the thermodynamics of residue interactions, $H_{in}$. The purpose of the growth constraint Hamiltonian, $H_{gc}$, is to prevent unrealistic geometries from occurring as the lattice is being constructed, such as the protein folding onto itself. This is prevented by assigning penalty terms that forbid any two consecutive residues from being placed along the same axis. Chirality is another natural and necessary physical property of peptides and the term $H_{ch}$ also applies penalties to preserve it. The more intricate of the three is the interaction Hamiltonian $H_{in}$. This term considers the 1st nearest neighbors (1-NN), but can be modified to go include up to 2nd nearest neighbors and beyond. This function incorporates the statistical potentials mentioned previously from \cite{MiyazawaJernigan}, which give the algorithm the ability to work with a full alphabet of amino acids, accounting for all 400 types of unique interactions (there are 20 canonical amino acids, capable of interacting with all 20 types). 
The algorithm can produce a model for the backbone alone, but also has the ability to include sidechains (also reduced to a single coarse grain sphere). For a more thorough explanation of the algorithm's Hamiltonian, we refer the readers to the original paper \cite{PFIvano}, its supplementary information, and associated github repository \footnote{https://github.com/qiskit-community/qiskit-research/tree/main/qiskit\_research/protein\_folding.}. There have been several efforts to generate a lattice-model for this application, using quantum algorithms \cite{perdomo1, perdomo2, babbush, babej2018coarsegrained, PFIvano, nathanwiebe_pf}. This algorithm \cite{PFIvano} could in theory be one of the more promising to draw inspiration from, as it produces 3D coordinates, incorporates a more realistic thermodynamic description, and is resource efficient. In terms of quantum resources required to create the backbone of the protein on a 3D lattice, \cite{PFIvano} scales quite well, as seen in the original publication and in our own resource estimates (see Figure 9a of main text). As a result, we adopted the algorithm and created an end-to-end workflow (see Figure 7 of main text) that allows us to compare the 3D predictions with experimental structures.

The algorithm takes the Hamiltonian discussed above and then produces an Ising Hamiltonian with $2^N$ terms (N being the number of amino acids in the input sequence), with each term and coefficient corresponding to an enumerated conformer. This is the first step - generating the conformational space, with each conformer corresponding to a specific eigen state and eigen value in the Ising hamiltonian. The subsequent step is to sample the space and find the minimum eigenvalue (i.e. the term corresponding to the lowest conformational energy). For small sequences, a classical exhaustive search is feasible, but for large values of \textit{N}, heuristic approaches become the only practical solution. Like many NISQ era algorithms, minimizing the hamiltonian is handled with a variational quantum eigensolver (VQE) \cite{BarkoutsosCVar}, a well-suited hybrid quantum-classical algorithm for this task. There are also alternatives that can be applied, including QAOA. At any rate, the quantum component handles the complex superposition states and entanglement inherent to quantum systems, while the classical component effectively optimizes the parameters of the quantum state, guiding the search for the lowest energy state. Moreover, by using a parameterized ansatz to prepare the quantum state, VQE allows an efficient exploration of the high-dimensional energy landscape inherent to protein structure prediction.

At this point, the algorithm has theoretically produced a conformation that should be reasonably close to the true global minimum. The resulting structure is stored in a generic XYZ file, with each set of coordinates corresponding to each residue's alpha carbon (C$\alpha$) position. Despite this rudimentary representation, producing this initial structure is by far the most challenging part, thus why it is handled by a quantum algorithm. To further enhance the utility of this prediction, a series of post processing steps must take place. Figure 7 (main text) illustrates these subsequent series of steps in the workflow, which can be efficiently handled classically. Classical algorithms are employed to first convert this XYZ file into a properly formatted C$\alpha$ trace pdb file \cite{xyz2pdb} (step 2). The all-atom structure must then be rebuilt, adding the missing carbonyl and amine groups of the backbone, and producing the sidechains using a program such as PULCHRA \cite{pulchra} (step 3). The missing backbone atoms are inserted in 2D fragments along the original axis in between the $\alpha$ carbons, ensuring the preservation of the original geometry predicted by the quantum algorithm. The last step is a final refinement of the all-atom model. This involves protonating the structure (adding hydrogen atoms at a chosen pH) and further minimizing these final set of coordinates with a steepest descent approach using a molecular mechanics force field. 

\subsection{Quantum circuit transpilation and optimization}\label{sec:resource_est}
When running circuits on IBM's quantum hardware, it is important to estimate the circuit depth. Since single qubits gates mostly have high gate fidelity, it is useful to analyze the ECR-depth for the circuits. Both of these gates are entangling gates that are common sources of the noise in the devices. An echoed cross-resonance gate (ECR) is equivalent to a CNOT gate up to single qubit rotations. We analyze the total number of ECRs and ECR-depth required in the circuit for different levels of optimization that are available in IBM's Qiskit package. The logical circuit created is transpiled first to match the topology and layout of the quantum processing unit. In this transpilation step, Qiskit Runtime \footnote{Find more details here https://www.ibm.com/quantum/qiskit-runtime.} has several default and custom options to optimize the circuit before executing on the hardware. To briefly describe, optimization level 0 maps the circuit to hardware layout with any explicit optimization. The higher levels of optimizations 1, 2 and 3 apply some optimization techniques such as adjacent gate collapsing, initial optimal mapping to physical qubits, noise-aware mapping. The higher optimization levels apply more sophisticated techniques that can potentially increase the circuit depth, but optimizes the mapping with respect to readout errors and gate fidelity. These optimization levels work really well in terms of decreasing overall circuit depth and ECR depth.

As we apply higher level optimizations, the total gate count and the ECR count decreases significantly. For example, with optimization level 0, the ECR depth is 618 for 20 amino acids. However, after applying optimization level 3, we see that it decreases to 100. Despite this immense gain in the circuit depth, we also observe that there is a sudden increase in both total number of ECR gates and the ECR depth as amino acid count reaches 22. This is due to the introduction of SWAP networks in the circuit to accommodate long-range connectivity among the physical qubits. In the unoptimized circuit, the relation seems to be quadratic while for different levels of optimization (as adopted in the Qiskit transpiler), the relation becomes quartic. In all instances, best fitting curves have an $R^2$ value larger than $0.9$. However, it would be interesting to see if one can apply a different circuit optimization technique to reduce the ECR-depth for larger proteins.

\subsection{Complexities in quadratizing the Hamiltonian}\label{sec:hamiltonian}
\begin{figure}[h!]
    \centering
    \includegraphics[width=16cm]{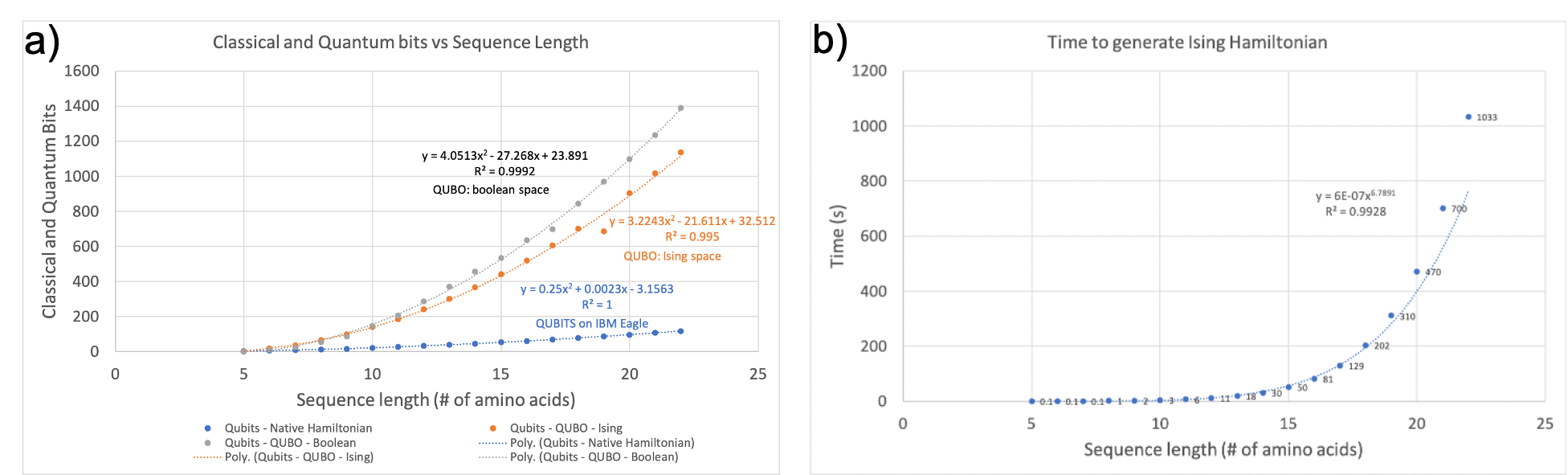}
    \caption{\textit{a) The qubit scaling as a function of sequence length (protein size) varies depending on the Hamiltonian form. The native Hamiltonian, which was designed to be solved on IBM's gate-based quantum hardware, scale quite efficiently (blue curve), with structures up to 22 amino acids able to be predicted with 127 qubits using our proposed workflow. As a comparison, the alternative would be to convert this native Hamiltonian into a QUBO Hamiltonian (in either Boolean or Ising space), which would allow the use of a quantum annealer. For all 3 cases, a quadratic regression provides a reasonable fit to the data, but the slopes are quite different, with a much slower growth in qubit requirement for the native Hamiltonian. b) While total run time is not measured for each case, we provide an estimate of the run times to just generate the native Ising Hamiltonian. It is largely exponential with respect to sequence length.}}
    \label{qubits}
\end{figure}

For performance comparison, we have also converted the Hamiltonian into a QUBO format and compared the resources needed. For this, the native Hamiltonian generated for each protein size was quadratized beforehand using an algorithm designed for this application \cite{fujitsu}, with the idea being that lower order QUBO Hamiltonians can be more feasibly handled by classical solvers or a quantum annealer. The program \footnote{https://github.com/flacrypto/compressed-quadratization.} quadratizes the native higher order Ising Hamiltonian efficiently (after processing it using an in house script we developed\footnote{https://github.com/jaidevjoshi83/hamiltonian\_format\_converter.git.}), but this comes at an expense. Specifically, reducing these higher order Hamiltonians to at most quadratic and linear terms requires the addition of penalties, which result in both ancillary qubits and additional terms in the Hamiltonian. The resulting quadratized Hamiltonians can be produced in either Ising or Boolean space, each with distinct growth rates as a function of increasing number of amino acids N. Figure \ref{qubits}a illustrates our estimates for the qubit scaling as a function of N. All 3 data sets are satisfied with a quadratic fit, but what is remarkable is the difference in slopes. It is rather clear that this algorithm \cite{PFIvano} is much more efficiently run directly on a gate-based quantum computer (blue line), with the coefficient of the $x^2$ term (0.25) being at least 12 times smaller than the other two. For the first 7 amino acids, the resources required in all 3 cases is comparable, but there is a quick divergence observed as the sequence grows. Additionally, it is not clear whether or not quadratization might benefit classical approaches. As a baseline, we used exhaustive search to determine the “ground truth” solutions. For 12 amino acids, the Hamiltonian \cite{PFIvano} requires 33 qubits per conformer, leading to $2^{33}$ permutations = 8.6 x $10^{9}$ mapping the conformational space.  On a desktop computer (Intel i9-13900 processor), a multi-threaded exhaustive search program required nearly one hour to run.  However the practical limit of exhaustive is perhaps 14 or 15 amino acids.  Heuristic search methods can extend the practical limit to a few more amino acids, while increasingly unable to guarantee finding the global minimum.  These estimates highlight the possibility of quantum advantage in tackling these problems. Furthermore, in agreement with Levinthal's paradox (as plotted in Figure2a of the main text), we found that the time for the algorithm to just generate the native Ising Hamiltonian as a function of sequence size is also exponential, with a sharp increase after 20 amino acids (see Figure \ref{qubits}b).

\subsection{Upper limit on quantum measurements}
\label{sec:measurement-formula}
We base our calculation on Section V of the supplemental section of \cite{kandala2017hardware}, and follow the same notation. For the Hamiltonians used in this work, the number of sets, $A = 1$ since all the terms are either constants or strings of Pauli Z and identity matrices. For the same reason, the number of Pauli terms, $T$ is same as $s_{\text{max}}$ which is the number of elements in the largest TPB set. Therefore, the error on the mean
energy $\langle H \rangle$ after taking $S$ samples for each Pauli operator is:

\begin{align}
    \epsilon &\le \sqrt{\frac{ h_{\text{max}}^2 \left(1 +  T\right)}{S} }
    \label{eq:ep1}
\end{align}

Now we solve Eq. ~\ref{eq:ep1} for the number of measurements, $S$.

\begin{align}
     S &\le \frac{ h_{\text{max}}^2 \left(1 +  T\right)}{\epsilon^2} 
    \label{eq:ep2}
\end{align}

% \break
\bibliographystyle{unsrt}
\bibliography{bibliography}